\documentclass[12pt,draftcls,onecolumn]{IEEEtran}
\usepackage{amsmath,graphicx}
\usepackage{epstopdf}
\usepackage{color}
\usepackage{wrapfig}
\usepackage{lipsum}
\usepackage{float}
\usepackage{colortbl,booktabs}
\usepackage{mathtools}
\usepackage[compress]{cite}
\definecolor{myRed}{rgb}{0.8, 0.2, 0.2}
\definecolor{myYellow}{rgb}{0.2,0.2,0.8}

\newcommand{\figref}[1]{Fig.~\ref{#1}}

\newcommand{\eqnref}[1]{(\ref{#1})}

\usepackage{amssymb}

\usepackage{multirow}
\usepackage{float}
 \usepackage{bm}
\usepackage{subfigure}
\usepackage{multicol}
\usepackage{algorithmic}
\usepackage{algorithm}
 \usepackage{mathrsfs}
\usepackage{bm}
\usepackage{amsthm}
\theoremstyle{definition}

\newtheorem{prop}{Proposition}
\theoremstyle{Proposition}
\theoremstyle{remark}
\newtheorem{remark}{Remark}
\theoremstyle{lemma}

\newtheorem{coro}{Corollary}
\theoremstyle{Corollary}
\usepackage{cite}
\usepackage{stfloats}%
\usepackage{lettrine}
\usepackage{enumitem}
\usepackage{setspace}
\usepackage{float}
\usepackage{graphicx}
\ifCLASSINFOpdf
\else
\fi
\begin{document}

\title{Splitting Receiver with Multiple Antennas}

\author{Yanyan Wang, \emph{Member, IEEE}, Wanchun Liu, \emph{Member, IEEE}, Xiangyun Zhou, \emph{Fellow, IEEE}\vspace{-0.65cm}}
\maketitle
\IEEEpeerreviewmaketitle

\begin{abstract}
Recently proposed splitting receivers, utilizing both coherently and non-coherently processed signals for detection, have demonstrated remarkable performance gain compared to conventional receivers in the single-antenna scenario.
In this paper, we propose a multi-antenna splitting receiver, where the received signal at each antenna is split into an envelope detection (ED) branch and a coherent detection (CD) branch, and the processed signals from both branches of all antennas are then jointly utilized for recovering the transmitted information.
We derive a closed-form approximation of the achievable mutual information (MI), in terms of the key receiver design parameters including the power splitting ratio at each antenna and the signal combining coefficients from all the ED and CD branches.
We further optimize these receiver design parameters and demonstrate important design insights for the proposed multi-antenna ED-CD splitting receiver:
1) the optimal splitting ratio is identical at each antenna, and 2)  the optimal combining coefficients for the ED and CD branches are the same, and each coefficient is proportional to the corresponding antenna’s channel power gain.
Our numerical results  also demonstrate the MI performance improvement of the proposed receiver over conventional non-splitting receivers.

\let\thefootnote\relax\footnote{
	Y. Wang is with School of Information Science and Technology, Southwest Jiaotong University, Chengdu, China  (email: yanyanwang@swjtu.edu.cn).
	W. Liu is with School of Electrical and Information Engineering, The University of
	Sydney, Australia (email: wanchun.liu@sydney.edu.au).
	X. Zhou is with School of Engineering, The Australian National University, Australia  (email: xiangyun.zhou@anu.edu.au). (\emph{Corresponding
author: Wanchun Liu}.)	
}
\end{abstract}

\section{Introduction}
Facing the demand for huge traffic, high date rate and  massive connectivity, many novel technologies~\cite{JiHLTZFL21}, e.g, intelligent reflecting surface (IRS)~\cite{WuZ20} and millimeter-wave massive multiple-input multiple-output (MIMO)~\cite{LiMSY21},  have been extensively studied.
The receiver design underlying these technologies  is  recognized as an essential element in the evolution of wireless communication systems. However, the basic receiver design principles behind each antenna  remain almost unchanged in the last few decades.

The conventional receiver schemes can be categorized into two classes:
coherent-detection (CD) receiver and non-coherent detection
receiver,  where  envelope detection (ED) and power detection (PD) are the most commonly-used detection mechanisms for non-coherent receivers~\cite{proakis2007digital, Jing2016Design, ElganiRPSGCR20}.
For the CD receiver, the received radio frequency (RF)-band signal is converted to a complex  (i.e., in-phase and quadrature) baseband signal by using a down-conversion circuit. Then the baseband signal is sampled and digitized through an analog-to-digital converter (ADC).
The CD receiver could detect the amplitude and phase of the received
signal~\cite{Zhou0H13}. For the ED and PD receivers,  the received RF-band signal is converted to a direct current signal by a rectifier and then digitized using an ADC. Such non-coherent receivers only detect the envelope or the power of the received signal~\cite{NawazSMPK21}.

Recently, a new receiver architecture was proposed in~\cite{Liu2017A} named as the  PD-CD splitting receiver, which introduced a novel joint coherent and non-coherent signal processing method.
For the splitting receiver architecture, the received signal is divided into two streams by a power splitter.  In~\cite{Liu2017A}, the two signal streams are first processed by the CD and PD circuits, respectively, and then jointly utilized for information detection.  The PD-CD splitting receiver achieves a higher date rate and lower symbol error rate than the conventional PD and CD receivers.
The follow-up work in~\cite{wang2020on} considered a more  practical noise model, where both the antenna noise and the processing (i.e., PD and CD) noises were taken into account.
In~\cite{WangLZ22}, another splitting receiver was proposed by replacing the PD circuit with an ED circuit for non-coherent signal processing. The ED-CD splitting receiver is more practical for wireless communications than the original one, since ED is the most commonly-used non-coherent detection mechanism.
The performance gain of the ED-CD splitting receiver over the conventional non-splitting benchmarks was established in terms of the achievable mutual information (MI).
These works in~\cite{wang2020on, WangLZ22} merely focused on the single-antenna scenario.

The splitting receiver architecture opens up an exciting research direction for the wireless communication systems, and we aim to investigate a multi-antenna ED-CD splitting receiver in the presence of both antenna and processing noises.
Multi-antenna receivers have been extensively utilized in the wireless communication systems due to its potential for providing high channel capacity and reliable communications~\cite{GoldsmithJJV03}.
Existing multi-antenna receivers often adopt CD for signal processing.
It remains largely unknown how to design an optimal multi-antenna splitting receiver and what the performance gain is. In particular, the optimal design needs to jointly determine the power splitting ratio at each antenna and the signal combining scheme for all the ED and CD branch signals, introducing new design challenges.

The main contributions of this paper are summarized as follows:
\begin{itemize}
  \item We establish a multi-antenna ED-CD splitting receiver architecture in the presence of both the antenna noise and processing noises. To understand the performance of the proposed receiver, we analytically characterize the achievable MI using a closed-form approximation. Simulation results show that the approximation is accurate at moderate and high signal-to-noise ratios (SNRs).
  \item We formulate an MI maximization problem in order to optimally design the key parameters of the proposed splitting receiver. In particular, we jointly optimize the splitting ratios and combining coefficients and present the solution in closed-form expressions. We show that the optimal power splitting ratio is identical at each antenna. Interestingly, the optimal signal combining coefficients for the ED and CD branches are the same and each coefficient is proportional to the corresponding antenna's channel power gain, which is different from the well-known maximum ratio combining (MRC) scheme for conventional multi-antenna receivers.
  \item Based on the optimal splitting ratios and combining coefficients, we investigate the achievable MI gain of the multi-antenna ED-CD splitting receiver over the conventional non-splitting receivers. Numerical results show  a notable performance improvement can be achieved under certain conditions as compared to the conventional receivers.
\end{itemize}

The remainder of the paper is organized as follows. Section II introduces the mathematical model of the proposed multi-antenna ED-CD splitting receiver. Section III analyzes the achievable MI performance. Section IV develops the problem formulation for jointly optimizing the splitting ratios and combining coefficients. The achievable MI  gain is investigated in Section V. The numerical results are presented in Section VI. Finally, we conclude the paper in Section VII.

\textit{Notation}: $\tilde{\cdot}$ and $|\cdot|$ denote a complex number and the absolute-value norm of a complex number, respectively.  $\mathcal{H}
(\cdot)$, $\mathcal{H}(\cdot, \cdot)$, $\mathcal{H}(\cdot|\cdot)$
represent the differential entropy, joint conditional differential
entropy and conditional differential
entropy, respectively. $\mathcal{I} (\cdot; \cdot)$ denotes the MI. $(\cdot)_r$ and $(\cdot)_i$ denote the real part and imaginary part of a complex number, respectively. $\mathbb{C}$ and $\mathbb{R}$  are the complex number and real number, respectively. In addition, $\mathcal{N}(m,\sigma^2)$ and $\mathcal{CN}(m,\sigma^2)$ denote the real-valued and complex-valued Gaussian distribution with mean $m$ and variance $\sigma^2$. $\mathbb{E}(\cdot)$ and $\textrm{Var}(\cdot)$ denote the expectation and variance of a random variable, respectively.
\section{Receiver Model}
\begin{figure}[t]
\centering
\includegraphics[width=0.7 \linewidth]{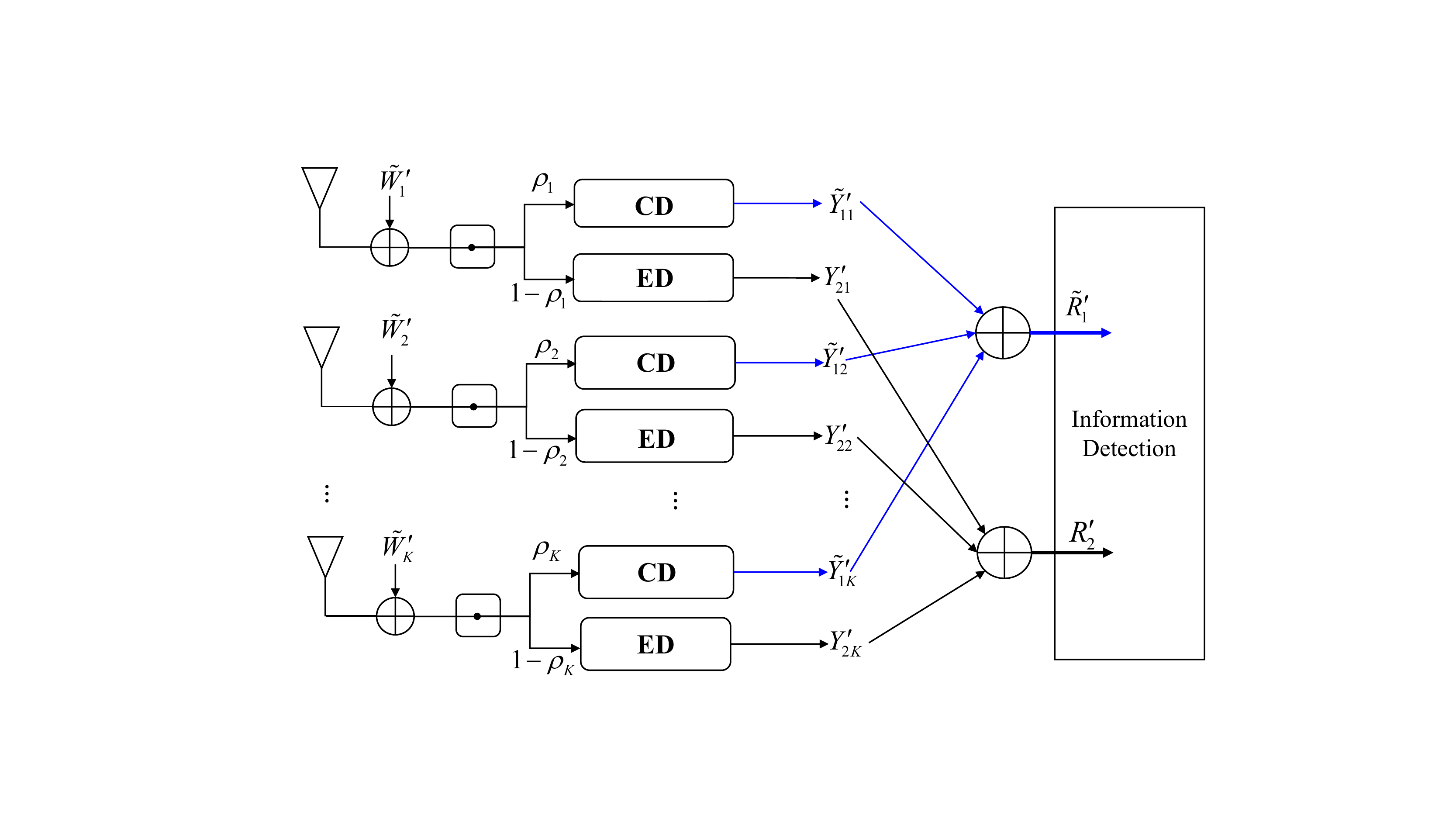}
\caption{$K$-antenna ED-CD splitting receiver architecture.}
\label{fig:Fig1sysmod1ed}
\end{figure}

We consider a  single-input multiple-output (SIMO) wireless communication system with a $K$-antenna ED-CD splitting receiver, as illustrated in \figref{fig:Fig1sysmod1ed}.

Let $ \tilde{X}$, $P$  and $\tilde{h}_k\triangleq |\tilde{h}_k|e^{j\phi_k}$ denote the transmitted signal, the average transmit power of the signal and the wireless channel coefficient for the $k$-th receive antenna, respectively. We assume that the channel state information is perfectly known at the receiver.
The received RF signal at the $k$-th antenna is split into two streams, one going into the CD processing branch and the other going into the ED processing branch with a power splitting ratio $\rho_k$, where $\rho_k \in [0,1]$. When $\rho_k=0$ (or $\rho_k=1$), the splitting receiver is degraded to the ED receiver (or the CD receiver).
Let ${\tilde{W^{'}_k}}\in \mathbb{C}$, $\tilde{Z}_k^{'}\in \mathbb{C}$, and $N_k\in \mathbb{R}$ denote  the antenna noise, the CD conversion noise and the ED rectifier noise, respectively, and the corresponding noise powers are denoted by $\sigma_{\textrm{A}}^2$, $\sigma_{\textrm{cov}}^2$ and $\sigma_{\textrm{rec}}^2$.
The received baseband CD and ED signals are
\begin{equation}
\label{equ:CDreceiverc}
{{{\tilde{Y}}}^{'}_{1k}} = \sqrt{\rho_k}(\sqrt{P}\tilde{h}_k{{\tilde{X}}}+{{\tilde{W^{'}_k}}})+{{\tilde{Z^{'}_k}}},
\end{equation}
\begin{equation}
\label{equ:EDreceiverc}
{{{Y}}^{'}_{2k}} = \sqrt{1-\rho_k}\big|\sqrt{P}\tilde{h}_k{{\tilde{X}}}+{{\tilde{W^{'}_k}}}\big|+{{N_k}}.
\end{equation}

For notation simplicity, we define ${{{\tilde{Y}}}_{1k}} \triangleq e^{-j\phi_k}{{{\tilde{Y}}}^{'}_{1k}}$, ${\tilde{W}_k}\triangleq e^{-j\phi_k}{{\tilde{W}_k}^{'}}$, and $\tilde{Z}_k\triangleq e^{-j\phi_k}{{\tilde{Z}^{'}_k}}$.
After linear scaling, the received signals can be represented as
\begin{equation}
\label{equ:CDreceiver1s}
{{{\tilde{Y}}}_{1k}} = {{\tilde{X}}}+\frac{{{\tilde{W}_k}}}{\sqrt{P}|\tilde{h}_k|}+\frac{{{\tilde{Z}_k}}}{\sqrt{\rho_k P}|\tilde{h}_k|},
\end{equation}
\begin{align}
\label{equ:EDreceiverchs1s}
{{{Y}}_{2k}} &= \bigg|{{\tilde{X}}}+\frac{{{\tilde{W}_k}}}{\sqrt{P}{|\tilde{h}_k|}}\bigg|+\frac{{{N_k}}}{\sqrt{(1-\rho_k)P}|\tilde{h}_k|}.
\end{align}

To recover the information carried by  ${\tilde{X}}$, we use a linear combining method to combine the CD branch signals (i.e., ${{{\tilde{Y}}}_{1k}}$,  $\forall k$) and ED branch signals (i.e., ${{{Y}}_{2k}}$,   $\forall k$), separately.
Let $\alpha_k$ and $\beta_k$ denote the combining coefficients for the ${{{\tilde{Y}}}_{1k}}$ and ${{{Y}}_{2k}}$, respectively.
Based on \eqref{equ:CDreceiver1s} and \eqref{equ:EDreceiverchs1s}, the combined CD and ED signals are given by
\begin{equation}
\label{equ:CDreceivermEDs}
{{{\tilde{R}}}_{1}} =\sum_{k=1}^{K} \alpha_k{{\tilde{X}}}+\sum_{k=1}^{K} \alpha_k\frac{{{\tilde{W}_k}}}{\sqrt{P}|\tilde{h}_k|}+\sum_{k=1}^{K} \alpha_k\frac{{{\tilde{Z}_k}}}{\sqrt{\rho_k P}|\tilde{h}_k|},
\end{equation}
\begin{equation}
\label{equ:PDreceiverchmEDs}
{{{R}}_{2}} = \sum_{k=1}^{K} \beta_k \bigg|{{\tilde{X}}}+\frac{{{\tilde{W}_k}}}{\sqrt{P}{|\tilde{h}_k|}}\bigg|+\sum_{k=1}^{K} \beta_k\frac{{{N_k}}}{\sqrt{(1-\rho_k)P}|\tilde{h}_k|}.
\end{equation}

Assuming that the CD branches of all $K$ antennas are the same, hence the conversion noises in all CD branches follow identical statistical distributions and independent from each other. Similarly, we assume that the rectifier noises in all ED branches follow independent and identical distributions. According to~\cite{WangLZ22} (and the references therein), we model the CD conversion noise as $\tilde{Z}\sim \mathcal{CN}(0,\sigma_{\textrm{cov}}^2)$ and the ED rectifier noise as $N\sim \mathcal{N}(0,\sigma_{\textrm{rec}}^2)$. Hence,
\eqref{equ:CDreceivermEDs} and \eqref{equ:PDreceiverchmEDs} can be
rewritten as
\begin{equation}
\label{equ:CDreceivermED1s}
{{{\tilde{R}}}_{1}} =\sum_{k=1}^{K} \alpha_k{{\tilde{X}}}+\sum_{k=1}^{K} \alpha_k\frac{{{\tilde{W}_k}}}{\sqrt{P}|\tilde{h}_k|}+\sqrt{\sum_{k=1}^{K} \frac{\alpha^2_k}{\rho_k P |\tilde{h}_k|^2}}{{\tilde{Z}}},
\end{equation}

\begin{align}
\label{equ:PDreceiverchmED1s}
&{{{R}}_{2}} = \sum_{k=1}^{K} \beta_k \bigg|{{\tilde{X}}}+\frac{{{\tilde{W}_k}}}{\sqrt{P}{|\tilde{h}_k|}}\bigg|+\!\!\sqrt{\sum_{k=1}^{K} \frac{\beta^2_k}{(1\!\!-\!\!\rho_k ) P |\tilde{h}_k|^2}}{N}.
\end{align}

From \eqnref{equ:CDreceivermED1s} and \eqnref{equ:PDreceiverchmED1s}, the two-dimensional (complex) signal ${{\tilde{R}}}_{1}$ and the one-dimensional (real) signal ${{R}}_{2}$ form an equivalent three-dimensional received signal $({{\tilde{R}}}_{1},{{R}}_{2})$. In the following, we investigate the MI between ${\tilde{X}}$ and $({{\tilde{R}}}_{1},{{R}}_{2})$, i.e., the amount of the transmitted information that can be recovered from the received signals.

\section{Mutual Information Analysis}

From \eqnref{equ:CDreceivermED1s} and \eqnref{equ:PDreceiverchmED1s}, the MI between the input signal ${{\tilde{X}}}$  and the output signal $({{{\tilde{R}}}_1}, {{{R}}_2})$  is expressed as
\begin{align}
\label{equ:MISPreceiverm}
\mathcal{I}\left({{\tilde{X}}};{{{\tilde{R}}}_1},{{{R}}_2}\right) &=\mathcal{H}\left({{{\tilde{R}}}_1},{{{R}}_2}\right)-\mathcal{H}\left({{{\tilde{R}}}_1},{{{R}}_2}\big|{{\tilde{X}}}\right)\notag\\
&=-\int_{{{{R}}_2}}\int_{{{{\tilde{R}}}_1}}f_{{{{\tilde{R}}}_1},{{{R}}_2}}({{{\tilde{r}}}_1},{{{r}}_2})\log_2\left(f_{{{{\tilde{R}}}_1},{{{R}}_2}}({{{\tilde{r}}}_1},{{{r}}_2})\right) \,\mathrm{d}{{{\tilde{r}}}_1}\mathrm{d}{{{r}}_2}\notag\\
&\,\,\,\,\,\,\,\,+\int_{{{\tilde{X}}}}\int_{{{\tilde{R}_1}}}\int_{{{{R_2}}}}f_{{{\tilde{X}}}}(\tilde{x})f_{{{\tilde{R}}_1},{{{R_2}}}}(\tilde{r}_1,{r_2}|\tilde{x})\log_2 \left(f_{{{\tilde{R}_1}},{{{R_2}}}}(\tilde{r}_1,{r_2}|\tilde{x})\right)\,\mathrm{d}{r_2}\mathrm{d}\tilde{r}_1\mathrm{d}\tilde{x},
\end{align}
where $f_{{{\tilde{X}}}}(\tilde{x})$ is the probability density
function (PDF) of the normalized input signal. The joint PDF of $({{{\tilde{R}}}_1},{{{R}}_2})$ is given by
\begin{align}
\label{equ:pdfy1y2}
f_{{{{\tilde{R}}}_1},{{{R}}_2}}({{{\tilde{r}}}_1},{{{r}}_2})&=\int_{\tilde{X}}\int_{\tilde{W}}f_{{{{\tilde{R}}}_1},{{{R}}_2}}({{{\tilde{r}}}_1},{{{r}}_2}|\tilde{x},\tilde{w})
f_{{{\tilde{X}}}}(\tilde{x})f_{{{\tilde{W}}}}(\tilde{w})\mathrm{d}{\tilde{w}}\mathrm{d}{\tilde{x}}\notag\\
&=\!\!\int_{\tilde{X}}\int_{\tilde{W}}\!\!f_{{{{\tilde{R}}}_1}}({{{\tilde{r}}}_1}|\tilde{x},\tilde{w})f_{{{{R}}_2}}({{{r}}_2}|\tilde{x},\tilde{w})
f_{{{\tilde{X}}}}(\tilde{x})
f_{{{\tilde{W}}}}(\tilde{w})\mathrm{d}{\tilde{w}}\mathrm{d}{\tilde{x}},
\end{align}
and the conditional joint PDF $f_{{{\tilde{R}_1}},{{{R_2}}}}(\tilde{r}_1,{r_2}|\tilde{x})$ is given by
\begin{align}
\label{equ:pdfy1y2co}
&f_{{{\tilde{R}_1}},{{{R_2}}}}(\tilde{r}_1,{r_2}|\tilde{x})=\!\!\int_{\tilde{W}}\!\!f_{{{{\tilde{R}}}_1}}({{{\tilde{r}}}_1}|\tilde{x},\!\tilde{w})f_{{{{R}}_2}}({{{r}}_2}|\tilde{x},\!\tilde{w})
f_{{{\tilde{W}}}}(\tilde{w})\,\mathrm{d}{\tilde{w}},
\end{align}
where the conditional PDFs $f_{{{{\tilde{R}}}_1}}({{{\tilde{r}}}_1}|\tilde{x},\tilde{w})$ and $f_{{{{R}}_2}}({{{r}}_2}|\tilde{x},\tilde{w})$ are $\mathcal{CN}\big(\sum_{k=1}^{K} \alpha_k{{\tilde{x}}}+\sum_{k=1}^{K} \alpha_k\frac{{{\tilde{w}_k}}}{\sqrt{P}|\tilde{h}_k|},\\ \sum_{k=1}^{K} \frac{\alpha^2_k}{\rho_k P |\tilde{h}_k|^2}\sigma_{\textrm{cov}}^2\big)$ and $\mathcal{N}\,\big( \sum_{k=1}^{K} \beta_k \big|{{\tilde{x}}}+\frac{{{\tilde{w}_{k}}}}{\sqrt{P}{|\tilde{h}_k|}}\big|,\sum_{k=1}^{K} \frac{\beta^2_k}{(1-\rho_k ) P |\tilde{h}_k|^2}\sigma_{\textrm{rec}}^2\big)$, respectively.

It is observed in \eqnref{equ:MISPreceiverm} that we need to calculate seven integrals for  evaluating the MI, which results in an extremely high computational complexity. Thus, we aim to derive the approximated MI in the high SNR regime. In addition, different choices of the distribution of $\tilde{X}$ also result in different MI and finding the optimal input distribution is extremely challenging if not impossible. Hence, motivated by the fact that the Gaussian distribution is the optimal input distribution for the conventional CD receivers, in this work, we adopt the assumption that $\tilde{X}\sim \mathcal{CN}(0, 1)$.

For MI approximation, we firstly define two noise variables ${{W}_{||k}}$ and  ${{W}_{\perp k}}$ as the projection of ${\tilde{W}_{k}}$ onto the same direction and the vertical direction of ${{\tilde{X}}}$, respectively. Note that ${{W}_{||k}}$ and ${{W}_{\perp k}}$  both follow zero-mean real Gaussian distributions with variance $\frac{\sigma_{\textrm{A}}^2}{2}$.
Then, \eqnref{equ:PDreceiverchmED1s} can be rewritten as
\begin{align}
\label{equ:EDreceiverchs1}
{{{R}}_{2}} = \sum_{k=1}^{K} \beta_k \bigg|{{\tilde{X}}}+\frac{{{{W}_{||k}}}}{\sqrt{P}{|\tilde{h}_k|}}+\frac{{{{W}_{\perp k}}}}{\sqrt{P}{|\tilde{h}_k|}}\bigg|+\sqrt{\sum_{k=1}^{K} \frac{\beta^2_k}{(1-\rho_k ) P |\tilde{h}_k|^2}}{N}.
\end{align}

Since ${{\tilde{X}}}+\frac{{{{W}_{||k}}}}{\sqrt{P}{|\tilde{h}_k|}}$ is vertical to $\frac{{{{W}_{\perp k}}}}{\sqrt{P}{|\tilde{h}_k|}}$ and the latter is much smaller than the former when $P$ is large,  we have the approximation $\bigg|{{\tilde{X}}}+\frac{{{{W}_{||k}}}}{\sqrt{P}{|\tilde{h}_k|}}+\frac{{{{W}_{\perp k}}}}{\sqrt{P}{|\tilde{h}_k|}}\bigg|\approx \bigg|{{\tilde{X}}}+\frac{{{{W}_{||k}}}}{\sqrt{P}{|\tilde{h}_k|}}\bigg| =\big|{{\tilde{X}}}\big|+\frac{{{{W}_{||k}}}}{\sqrt{P}{|\tilde{h}_k|}}$ in the high SNR regime.
Then, the combined ED-branch signal in \eqnref{equ:EDreceiverchs1} is approximated as
\begin{align}
\label{equ:PDreceiverchmED1}
&{{{R}}_{2}} \approx \sum_{k=1}^{K} \beta_k \big|{{\tilde{X}}}\big|\!\!+\!\!\sum_{k=1}^{K} \beta_k\frac{{{{W}_{||k}}}}{\sqrt{P}{|\tilde{h}_k|}}\!\!+\!\!\sqrt{\sum_{k=1}^{K} \frac{\beta^2_k}{(1\!\!-\!\!\rho_k ) P |\tilde{h}_k|^2}}{N}.
\end{align}
Based on \eqnref{equ:CDreceivermED1s} and \eqnref{equ:PDreceiverchmED1}, we obtain the approximated MI as below.
\begin{prop}
In the high SNR regime, the achievable MI of the multi-antenna ED-CD splitting receiver with the power  splitting ratios $[\rho_1, \rho_2,\cdots,\rho_K] \in [0,1]^K\backslash \{\mathbf{0},\mathbf{1}\} $, and the combining coefficients $[\alpha_1,\alpha_2,\cdots,\alpha_K]$ and $[\beta_1,\beta_2,\cdots,\beta_K]$  can be approximated as
\begin{align}
\label{equ:MIendm}
\mathcal{I}({{{\tilde{X}}}};{{{\tilde{R}}}_1}, {{{R}}_2})\approx &\frac{1}{2}\log_2\left({A^2+1}\right)-\frac{1}{2}\log_2\left( {\sum_{k=1}^{K} B^2_k\sigma_{\textrm{A}}^2+C^2\sigma_{\textrm{cov}}^2}\right)-\notag\\
&\frac{1}{2}\log_2\left({{\sum_{k=1}^{K} \bigg(\frac{B_k}{\sqrt{1+A^2}}+\frac{B'_kA}{\sqrt{1+A^2}} \bigg)^2}\sigma_{\textrm{A}}^2+C^2\sigma_{\textrm{cov}}^2} \right).
\end{align}
where $A\triangleq \gamma \sum_{k=1}^{K} \beta_k$, $B_k\triangleq\frac{\alpha_k}{A'\sqrt{P}|\tilde{h}_k|}$, $C\triangleq\frac{\sqrt{\sum_{k=1}^{K} \frac{\alpha^2_k}{\rho_k P |\tilde{h}_k|^2}}}{A'}$, $A^{'}\triangleq \sum_{k=1}^{K} \alpha_k$,  $B^{'}_k\triangleq \frac{\gamma \beta_k}{\sqrt{P}{|\tilde{h}_k|}}$, $C^{'}\triangleq \gamma\sqrt{\sum_{k=1}^{K} \frac{\beta^2_k}{(1-\rho_k ) P |\tilde{h}_k|^2}}$,
and
\begin{align}
\label{equ:vagamma}
\gamma\triangleq\frac{{\sqrt{\sum_{k=1}^{K} \frac{\alpha^2_k}{\rho_k P |\tilde{h}_k|^2}}}\sigma_{\textrm{cov}}}{\sqrt{2}\sqrt{\sum_{k=1}^{K} \frac{\beta^2_k}{(1-\rho_k ) P |\tilde{h}_k|^2}}A'\sigma_{\textrm{rec}}}.
\end{align}

\emph{Proof}: See Appendix A.\hfill

\label{proposition:prop3}
\end{prop}

In Section VI, we will show that the approximation is accurate at moderate and high SNRs.  Hence, the obtained analytical approximation significantly reduces the computational complexity of the MI compared to \eqnref{equ:MISPreceiverm}.
\section{Joint Optimal Design of Splitting Ratios and Combining Coefficients}
Under the multi-antenna splitting receiver architecture, the key receiver design parameters are the power splitting ratios, i.e, $\boldsymbol{\rho}=[\rho_1,\rho_2,\cdots,\rho_K]$, the combining coefficients of all the CD branches, i.e., $\boldsymbol{\alpha}=[\alpha_1,\alpha_2,\cdots,\alpha_K]$, and the combining coefficients of all the ED branches, i.e., $\boldsymbol{\beta}=[\beta_1,\beta_2,\cdots,\beta_K]$. In this section, we propose the problem below to jointly optimize these design parameters for achieving the maximum MI:

\begin{equation}
\label{equ:MIopti}
  \begin{aligned}
      & \max_{\boldsymbol{\rho},\boldsymbol{\alpha},\boldsymbol{\beta}}\,\, \mathcal{I}({{{\tilde{X}}}};{{{\tilde{R}}}_1}, {{{R}}_2})\\
        s.t. &\begin{array}{r@{\quad}r@{}l@{\quad}l}
                          &  \boldsymbol{\rho}\in [0,1]^K\backslash \{\mathbf{0},\mathbf{1}\}.
                           \end{array}
  \end{aligned}
\end{equation}

The following proposition presents the optimal solution of problem \eqnref{equ:MIopti} and the maximum achievable MI.
\begin{prop}
\label{proposition:prop4}
 In the high SNR regime, the followings hold:
 \begin{enumerate}[label=(\roman*)]
\item  the maximum achievable MI is
 \begin{align}
\label{equ:MIoptimsp1}
\mathcal{I}_{\textrm{max}}({{{\tilde{X}}}};{{{\tilde{R}}}_1}, {{{R}}_2})& \approx \log_2\bigg(\sum_{k=1}^{K}P|\tilde{h}_k|^2\bigg)\notag\\
&\,\,\,\,-\frac{1}{2}\log_2\bigg(\bigg({{\frac{\sigma_{\textrm{cov}}^2}{\rho^*  }}}+\sigma_{\textrm{A}}^2 \bigg)
\bigg({\sigma_{\textrm{A}}^2} +\frac{{{\frac{\sigma_{\textrm{cov}}^2}{\rho^*}}}2 \frac{\sigma_{\textrm{rec}}^2}{(1-\rho^* ) }}{{{\frac{\sigma_{\textrm{cov}}^2}{\rho^*}}}+2 \frac{\sigma_{\textrm{rec}}^2}{(1-\rho^* ) }}\bigg)\bigg),
\end{align}
where $\rho^*$ is given by
\begin{align}
\label{equ:oprhosp1-2}
\rho^*=
\begin{cases}
\Upsilon, & \sigma_{\textrm{{cov}}}^2> 4 \sigma_{\textrm{{rec}}}^2,\\
1, & \textrm{else},
 \end{cases}
\end{align}
and $\Upsilon$ is given by
\begin{align}
\label{equ:MIendsim22-11}
\Upsilon=\frac{\sigma_{\textrm{cov}}^2 (\sigma_{\textrm{cov}}^2 - 2 \sigma_{\textrm{rec}}^2) ( \sigma_{\textrm{A}}^2 + 2\sigma_{\textrm{rec}}^2) -
\sqrt{2
  \Psi}}{\sigma_{\textrm{A}}^2 (\sigma_{\textrm{cov}}^2 - 4 \sigma_{\textrm{rec}}^2) (\sigma_{\textrm{cov}}^2 - 2 \sigma_{\textrm{rec}}^2)},
\end{align}
where $\Psi=\sigma_{\textrm{cov}}^4 (\sigma_{\textrm{A}}^2 + \sigma_{\textrm{cov}}^2 - 2 \sigma_{\textrm{rec}}^2) (\sigma_{\textrm{cov}}^2 - 2 \sigma_{\textrm{rec}}^2) \sigma_{\textrm{rec}}^2 (\sigma_{\textrm{A}}^2 +2
     \sigma_{\textrm{rec}}^2)$.
\item  the  optimal power splitting ratio at each antenna is identical  and equals  $\rho^*$ in \eqnref{equ:oprhosp1-2}.
\item  the optimal combining coefficients satisfy the following properties
 \begin{equation}
 \label{equ:optCC1}
 \begin{split}
     & \alpha^*_k=c_{\alpha}|\tilde{h}_k|^2, \\
      & \beta^*_k=c_{\beta}|\tilde{h}_k|^2,
 \end{split}
 \end{equation}
where $k=1,2,\cdots K$. $c_{\alpha}$ and $c_{\beta}$ are two arbitrary non-zero constants.
\end{enumerate}

 \emph{Proof}: See Appendix B.\hfill
\end{prop}
\begin{remark}
The optimal combining coefficients given in~\eqnref{equ:optCC1} are different from the widely-known equal gain combining (EGC) and MRC used in the conventional multi-antenna receivers\footnote[1]{It is well-known that the MRC scheme is optimal for multi-antenna CD receivers. Based on the similar analytical steps, it can be readily proved that MRC is also optimal for multi-antenna ED receivers. Note that in this work, we refer to MRC as the well-known optimal combining scheme for the conventional multi-antenna receivers.}. If one was to use EGC in the splitting receiver, we would have $\alpha_1=\alpha_2=\cdots=\alpha_K$ and $\beta_1=\beta_2=\cdots=\beta_K$. If one was to use MRC in the splitting receiver, we would have $\alpha_k = c_{\alpha} |\tilde{h}_k| $ and $\beta_k = c_{\beta} |\tilde{h}_k|$. In particular, the MRC scheme requires the combining coefficient to be linearly proportional to the magnitude of the corresponding channel. In contrast, the optimal combining scheme in~\eqnref{equ:optCC1} requires the combining coefficient to be linearly proportional to the power (i.e., magnitude squared) of the corresponding channel.
 \end{remark}

\begin{remark}
From the proof of Proposition  \ref{proposition:prop4}, one can readily derive the MI results when the EGC or MRC scheme is used for the splitting receiver. Note that for both EGC and MRC schemes, the combining coefficients satisfy
\begin{align}
\label{equ:optcoe}
 \frac{\alpha_1}{\beta_1}=\frac{\alpha_2}{\beta_2}=\cdots=\frac{\alpha_K}{\beta_K}.
 \end{align}

Following the proof of Proposition  \ref{proposition:prop4}, one can show that the optimal splitting ratio at each antenna is identical and equals $\rho^*$ in \eqnref{equ:oprhosp1-2} under the condition of~\eqnref{equ:optcoe}. Thus, we have the following MI results.
\end{remark}

\begin{coro}
\label{corollary:coro1}
When the EGC scheme is used for signal combining in the multi-antenna ED-CD splitting receiver, the achievable MI in the high SNR regime is given by
\begin{align}
\label{equ:MIoptegc}
\mathcal{I}_{\textrm{EGC}}({{{\tilde{X}}}};{{{\tilde{R}}}_1}, {{{R}}_2})&\approx\log_2\bigg(\frac{PK^2}{\sum_{k=1}^{K}\frac{1}{|\tilde{h}_k|^2}}\bigg)\notag\\
&-\frac{1}{2}\log_2\bigg(\bigg({{\frac{\sigma_{\textrm{cov}}^2}{\rho^*  }}}+\sigma_{\textrm{A}}^2 \bigg)
\bigg({\sigma_{\textrm{A}}^2} +\frac{{{\frac{\sigma_{\textrm{cov}}^2}{\rho^*}}}2 \frac{\sigma_{\textrm{rec}}^2}{(1-\rho^* ) }}{{{\frac{\sigma_{\textrm{cov}}^2}{\rho^*}}}+2 \frac{\sigma_{\textrm{rec}}^2}{(1-\rho^* ) }}\bigg)\bigg).
\end{align}
\end{coro}

\begin{coro}
\label{corollary:coro2}
When the MRC scheme is used for signal combining in the multi-antenna ED-CD splitting receiver, the achievable MI in the high SNR regime is given by
\begin{align}
\label{equ:MIoptmrc}
\mathcal{I}_{\textrm{MRC}}({{{\tilde{X}}}};{{{\tilde{R}}}_1}, {{{R}}_2})&\approx\log_2\left(\frac{P\bigg(\sum_{k=1}^{K}|\tilde{h}_k|\bigg)^2}{K}\right)\notag\\
&-\frac{1}{2}\log_2\bigg(\bigg({{\frac{\sigma_{\textrm{cov}}^2}{\rho^*  }}}+\sigma_{\textrm{A}}^2 \bigg)
\bigg({\sigma_{\textrm{A}}^2} +\frac{{{\frac{\sigma_{\textrm{cov}}^2}{\rho^*}}}2 \frac{\sigma_{\textrm{rec}}^2}{(1-\rho^* ) }}{{{\frac{\sigma_{\textrm{cov}}^2}{\rho^*}}}+2 \frac{\sigma_{\textrm{rec}}^2}{(1-\rho^* ) }}\bigg)\bigg).
\end{align}
\end{coro}

From \eqnref{equ:MIoptimsp1}, \eqnref{equ:MIoptegc}  and \eqnref{equ:MIoptmrc}, according to the mean value inequality, we have
\begin{align}
\label{equ:MIequ}
\frac{K^2}{\sum_{k=1}^{K}\frac{1}{|\tilde{h}_k|^2}}\leq\frac{\bigg(\sum_{k=1}^{K}|\tilde{h}_k|\bigg)^2}{K}\leq\sum_{k=1}^{K}|\tilde{h}_k|^2.
\end{align}

When $|\tilde{h}_1|=|\tilde{h}_2|=\cdots=|\tilde{h}_K|$, the equality holds. Therefore, the optimal combining coefficients  in \eqnref{equ:optCC1} is superior to the MRC and EGC schemes.
\section{Mutual Information Performance Gain}
In this section, based on the optimal splitting ratios and combining coefficients, we analyze the MI performance gain of the multi-antenna ED-CD splitting receiver as compared to the conventional CD receiver and ED receiver. In other words, we aim to quantify the increase in MI by using the splitting ratio as compared to the conventional receivers.

 The MI performance gain of the ED-CD splitting receiver is defined as
\begin{align}
\label{equ:MIgain2g}
&G_{\textrm{MI}}=  \mathcal{I}(\!{{\tilde{X}}};{{{\tilde{R}}}_1},\!{{{R}}_2}\!)|_{\boldsymbol{\rho}=\boldsymbol{\rho^*}, \boldsymbol{\alpha}=\boldsymbol{\alpha^*},\boldsymbol{\beta}=\boldsymbol{\beta^*}}
-\max\{\mathcal{I}(\!{{\tilde{X}}};{{{\tilde{R}}}_1},\!{{{R}}_2}\!)|_{\rho=1}, \mathcal{I}(\!{{\tilde{X}}};{{{\tilde{R}}}_1},\!{{{R}}_2}\!)|_{\rho=0}\},
\end{align}
where $\boldsymbol{\rho^*}$, $\boldsymbol{\alpha^*}$, and $\boldsymbol{\beta^*}$ can be achieved by  Proposition \ref{proposition:prop4}. In \eqnref{equ:MIgain2g}, the maximum achievable MI of the splitting receiver is given in \eqnref{equ:MIoptimsp1} and the MI of the CD receiver is
\begin{align}
\label{equ:MICD}
\mathcal{I}({{{\tilde{X}}}};{{{\tilde{R}}}_1}, {{{R}}_2})|_{\rho=1}=\log_2\left(1+\frac{\sum_{k=1}^{K} P|\tilde{h}_k|^2}{\sigma_{\textrm{cov}}^2+\sigma_{\textrm{A}}^2}\right).
\end{align}

Since $\mathcal{I}(\!{{\tilde{X}}};{{{\tilde{R}}}_1},\!{{{R}}_2}\!)|_{\rho=1}> \mathcal{I}(\!{{\tilde{X}}};{{{\tilde{R}}}_1},\!{{{R}}_2}\!)|_{\rho=0}$ in the high SNR regime~\cite{Zhou0H13}, the performance gain of the ED-CD splitting receiver is represented as
\begin{align}
\label{equ:MIgain2as}
\mathop{\lim} \limits_{P\rightarrow\infty}G_{\textrm{MI}}&= \mathcal{I}(\!{{\tilde{X}}};{{{\tilde{R}}}_1},\!{{{R}}_2}\!)|_{\rho=\rho^*}\!-\!\mathcal{I}(\!{{\tilde{X}}};{{{\tilde{R}}}_1},\!{{{R}}_2}\!)|_{\rho=1}\notag\\
&=\log_2\bigg(\sum_{k=1}^{K}P|\tilde{h}_k|^2\bigg)-\log_2\!\! \left(1+\frac{\sum_{k=1}^{K} P|\tilde{h}_k|^2}{\sigma_{\textrm{A}}^2+\sigma_{\textrm{cov}}^2}\right)\notag\\
&-\frac{1}{2}\log_2\bigg(\bigg({{\frac{\sigma_{\textrm{cov}}^2}{\rho^* }}}+\sigma_{\textrm{A}}^2 \bigg)
\bigg({\sigma_{\textrm{A}}^2} +\frac{{{\frac{\sigma_{\textrm{cov}}^2}{\rho^*}}}2 \frac{\sigma_{\textrm{rec}}^2}{(1-\rho^* ) }}{{{\frac{\sigma_{\textrm{cov}}^2}{\rho^*}}}+2 \frac{\sigma_{\textrm{rec}}^2}{(1-\rho^*) }}\bigg)\bigg).
\end{align}

When $P$ is large, $\frac{\sum_{k=1}^{K} P|\tilde{h}_k|^2}{\sigma_{\textrm{A}}^2+\sigma_{\textrm{cov}}^2}\gg1$. The first two terms of~\eqnref{equ:MIgain2as} are simplified as $\log_2(\sigma_{\textrm{A}}^2+\sigma_{\textrm{cov}}^2)$.
Thus, the MI gain of the multi-antenna ED-CD splitting receiver is approximated as
\begin{align}
\label{equ:MIgain2a}
&\mathop{\lim} \limits_{P\rightarrow\infty}G_{\textrm{MI}}\approx
\frac{1}{2}\log_2\left(\frac{\rho^{*} ((1-\rho^{*})\sigma_{\textrm{cov}}^2+2\rho^{*} \sigma_{\textrm{rec}}^2)(\sigma_{\textrm{A}}^2+\sigma_{\textrm{cov}}^2)^2}{(\rho^{*} \sigma_{\textrm{A}}^2+\sigma_{\textrm{cov}}^2)(2\rho^{*} \sigma_{\textrm{rec}}^2\sigma_{\textrm{A}}^2+(1-\rho^{*})\sigma_{\textrm{cov}}^2\sigma_{\textrm{A}}^2\!\!+\!\!2\sigma_{\textrm{cov}}^2\sigma_{\textrm{rec}}^2)} \right).
\end{align}

By analyzing the MI gain  in \eqnref{equ:MIgain2a}, we can derive the following result.
\begin{coro}
In the high SNR regime, the achievable  MI  gain of the multi-antenna ED-CD splitting receiver is given by
\begin{equation}
\label{equ:MIgain2}
\begin{aligned}
G_{\textrm{MI}}=
\begin{cases}
  \eqnref{equ:MIgain2a}, &  \sigma_{\textrm{{cov}}}^2> 4 \sigma_{\textrm{{rec}}}^2,\\
  0, &\textrm{else}.
\end{cases}
 \end{aligned}
\end{equation}
\label{corollary: coro3}
\end{coro}
From Corollary \ref{corollary: coro3}, when the optimal splitting ratio is $\rho^*=\Upsilon$,  the multi-antenna ED-CD splitting receiver provides higher MI than the conventional CD receiver ($\rho=1$), which is the same as proved for the single-antenna case~\cite{WangLZ22}, i.e., $\mathcal{I}(\!{{\tilde{X}}};{{{\tilde{R}}}_1},\!{{{R}}_2}\!)|_{\rho^*=\Upsilon}\!>\!\mathcal{I}(\!{{\tilde{X}}};{{{\tilde{R}}}_1},\!{{{R}}_2}\!)|_{\rho=1}$.

\begin{remark}
The multi-antenna ED-CD splitting receiver achieves higher MI than the conventional CD and ED receivers. As shown in Corollary \ref{corollary: coro3},  the obtained MI gain of the multi-antenna ED-CD splitting receiver approaches to a constant value (which can be shown to be larger than zero) in the high SNR regime when $\sigma_{\textrm{{cov}}}^2> 4 \sigma_{\textrm{{rec}}}^2$.
Also, this asymptotic performance gain  is independent of the number of antennas $K$ in high SNR.
\end{remark}
\section{Numerical Results}
In this section, we present numerical results to demonstrate the performance of the multi-antenna ED-CD splitting receiver.
In practice, the rectifier noise power $\sigma_{\textrm{rec}}^2$  and the antenna noise power $\sigma_{\textrm{A}}^2$ are much smaller than the conversion noise power $\sigma_{\textrm{cov}}^2$~\cite{WangLZ22}.
Thus, we set the noise powers $\sigma_{\textrm{cov}}^2=1,  \sigma_{\textrm{A}}^2=\sigma_{\textrm{rec}}^2=0.01$ in the following results.

\begin{figure}[t]
\centering
\includegraphics[width=0.7 \linewidth]{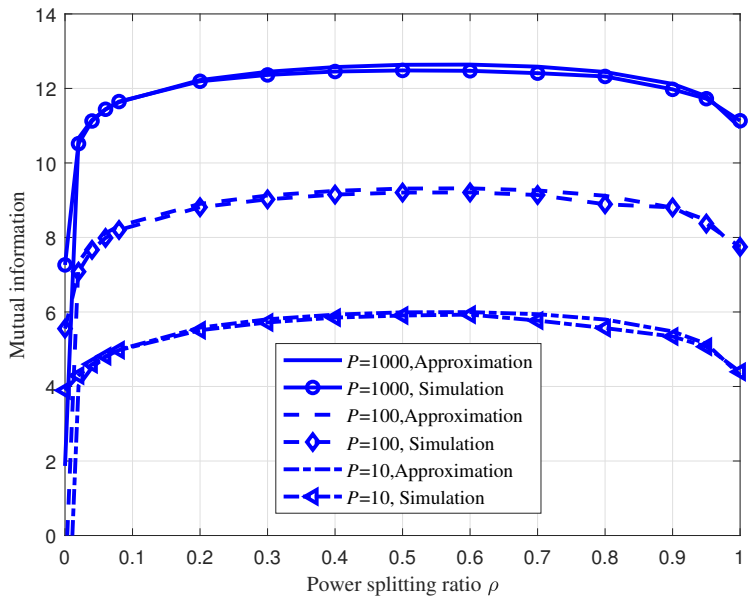}
\caption{MI versus the power splitting ratio $\rho$, $\alpha_1=\alpha_2=\beta_1=\beta_2=0.50$ and $|\tilde{h}_1|=|\tilde{h}_2|=1$.}
\label{fig:FigK2MI}
\end{figure}

In~\figref{fig:FigK2MI}, we plot the achievable MI given in \eqnref{equ:MIendm} and the simulated MI against the power splitting
ratio for different transmit power values. The simulated MI is obtained by the Monte Carlo based histogram method, which is commonly used for differential entropy estimation. This method gives accurate evaluation of the differential entropy (and hence the MI) when the bin width of the histogram is sufficiently small.
As shown in~\figref{fig:FigK2MI},  the analytical approximation is accurate as compared to the simulated MI in moderate and high SNRs (from $P=10$ to $1000$), which verifies the accuracy of the approximation expression. In addition, it is observed that the optimal splitting ratios are roughly the same for different signal powers. When $P=10$, $100$ and $1000$,  the optimal $\rho$ that maximizes the MI is about $0.56$. This observed result of optimal $\rho$ is consistent with the value calculated in \eqnref{equ:oprhosp1-2}.

\begin{figure}[H]
\centering
\includegraphics[width=0.7 \linewidth]{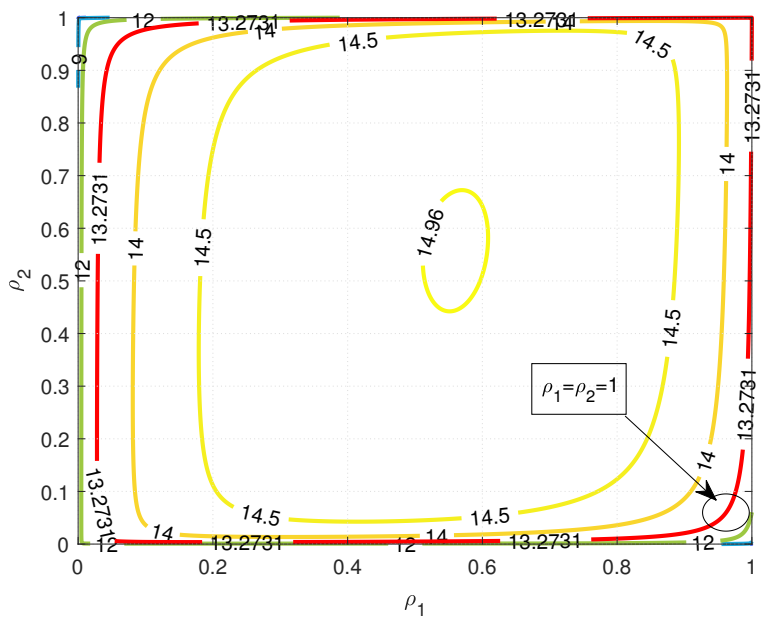}
\caption{MI versus the power splitting ratio with two antennas, $P=1000$, $|\tilde{h}_1|=1,|\tilde{h}_2|=3$, $\alpha_1=\beta_1=0.1, \alpha_2=\beta_2=0.9$.} 
\label{fig:FigK23D1}
\end{figure}

In \figref{fig:FigK23D1}, we plot the contour of the MI for different power splitting ratios with given combining coefficients. It can be observed that the MI varies significantly with different power splitting ratios.  When the power splitting ratios are equal for the two antennas and both are about $0.56$, the MI  obtains the maximum and is about $14.96$, which is superior to the conventional CD receiver with $\rho=1$.

\begin{figure}[t]
\centering
\includegraphics[width=0.7 \linewidth]{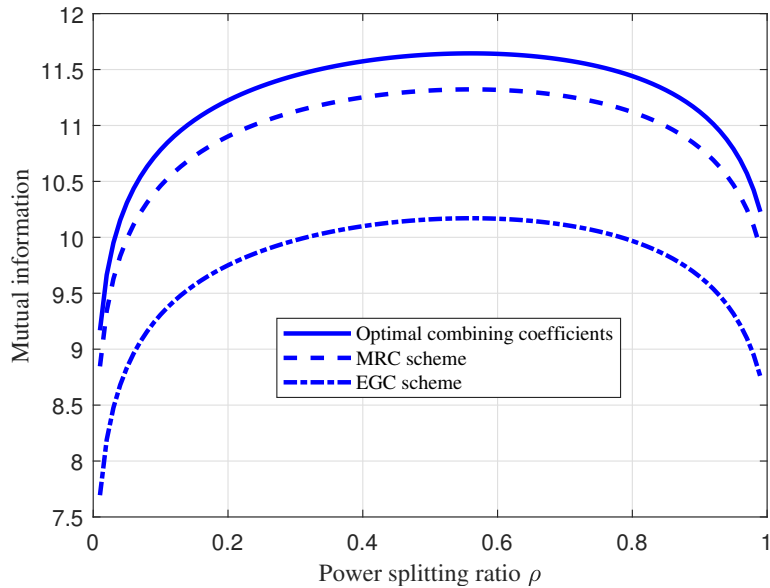}
\caption{MI versus the power splitting ratio with two antennas, $P=100$.} 
\label{fig:FigKMIdiffcom}
\end{figure}

\figref{fig:FigKMIdiffcom} shows the MI versus the power splitting ratio for different combining schemes. For this figure, the number of the receive antennas is 2 and the  corresponding channel magnitudes of the two antennas are respectively set as $|\tilde{h}_1|=1$ and $|\tilde{h}_2|=3$. It is observed in~\figref{fig:FigKMIdiffcom} that the achieved MI with optimal combining scheme outperforms the conventional EGC and MRC schemes, which verifies the superiority of the proposed combining method.

\begin{figure}[H]
\centering
\includegraphics[width=0.7 \linewidth]{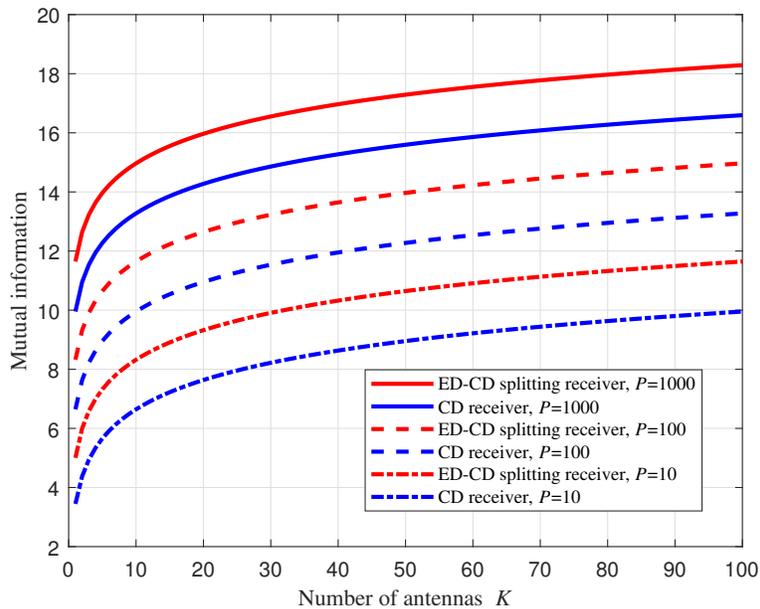}
\caption{MIs of the ED-CD splitting receiver and the CD
receiver versus the number of antennas $K$,  $|\tilde{h}_k|=1$ for all $k$.}
\label{fig:FigMISRvsMICD}
\end{figure}

In \figref{fig:FigMISRvsMICD}, we plot
the achievable MIs given in \eqnref{equ:MIoptimsp1} and \eqnref{equ:MICD} for the ED-CD  splitting receiver and the conventional CD receiver, respectively, against the number of antennas for different transmit power values. It is observed that as the signal power (from $P=10$ to $1000$) and the number of antennas (from $K=1$ to $100$)  increase, the obtained MI of the ED-CD splitting receiver is enhanced. Also, the ED-CD splitting receiver achieves higher MI than that of the CD receiver. For example, when $K=10$ and $P=100$, the MI gap between the ED-CD splitting receiver and the CD receiver is about 1.69, which is roughly $17\%$ difference.

\begin{figure}[H]
\centering
\includegraphics[width=0.7 \linewidth]{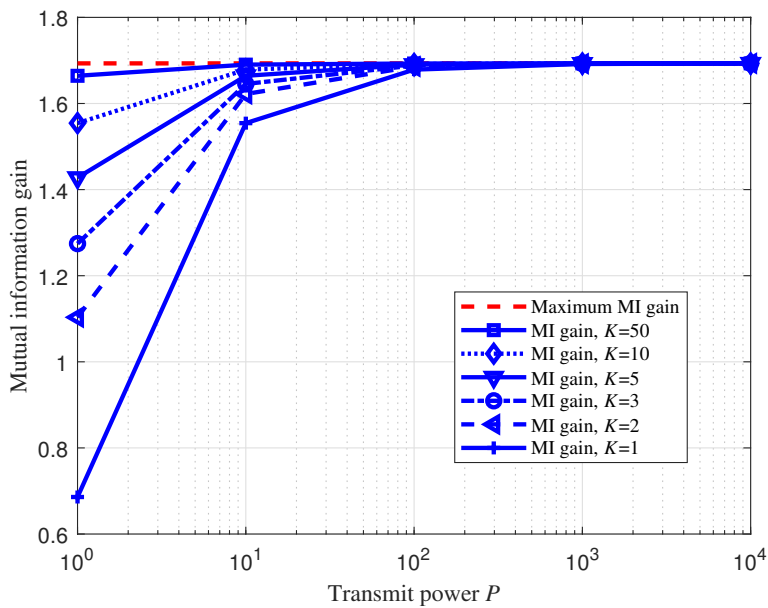}
\caption{MI gain versus the transmit power $P$ with different number of antennas, $|\tilde{h}_k|=1$ for all $k$.}
\label{fig:FigK2MIgain1}
\end{figure}

\figref{fig:FigK2MIgain1} depicts the MI gain defined in~\eqnref{equ:MIgain2as} versus the transmit power with different number of antennas. It is observed that when the signal power is below 100, the MI gain has not yet reached its high-SNR asymptotic value, hence, also increases as the number of antennas increases.
When the signal power is larger than 100, the MI gain reaches its asymptotic value and hence becomes independent of the number of antennas $K$.
Therefore, one can infer that the benefit of the proposed splitting receiver increases with the number of antennas at low-to-moderate SNRs.

\section{Conclusions}
In this paper, we have proposed the multi-antenna ED-CD splitting receiver architecture for the wireless communication system.
By characterizing the MI performance of the proposed receiver, we have obtained optimal design parameters, including the optimal splitting ratio at each antenna and the optimal combining coefficients for all the ED branches and CD branches.  In particular, the optimal combining coefficients are shown to be different from the existing knowledge of MRC for conventional multi-antenna receivers. Our numerical results have also demonstrated notable performance improvement of using the splitting receiver as compared to the conventional receivers.
\begin{appendices}
\section{ Proof of Proposition 1}
Due to the property of MI invariance under
scaling of random variables~\cite{BookInfo},  \eqref{equ:CDreceivermED1s} and \eqref{equ:PDreceiverchmED1} can be linearly scaled as
\setcounter{equation}{0}
\renewcommand\theequation{A.\arabic{equation}}
\begin{equation}
\label{equ:CDreceivermED2}
{{{\tilde{R}}}_{1}} ={{\tilde{X}}}+\frac{\sum_{k=1}^{K}\frac{ \alpha_k}{\sqrt{P}|\tilde{h}_k|}}{\sum_{k=1}^{K} \alpha_k}{\tilde{W}_k}+\frac{\sqrt{\sum_{k=1}^{K} \frac{\alpha^2_k}{\rho_k P |\tilde{h}_k|^2}}}{\sum_{k=1}^{K} \alpha_k}{{\tilde{Z}}},
\end{equation}
\begin{align}
\label{equ:PDreceiverchmED2}
{{{R}}_{2}} \!\!\approx \!\!\sum_{k=1}^{K}\! \gamma\! \beta_k \big|{{\tilde{X}}}\big|\!\!+\!\!\gamma\sum_{k=1}^{K} \beta_k\frac{{{{W}_{||k}}}}{\sqrt{P}{|\tilde{h}_k|}}\!\!+\!\!\gamma\!\!\sqrt{\!\!\sum_{k=1}^{K} \!\!\frac{\beta^2_k}{(1\!\!-\!\!\rho_k ) P |\tilde{h}_k|^2}}{N}.
\end{align}
By using  $A$, $B_k$, $C$, $A^{'}$, $B^{'}_k$ and $C^{'}$ defined under~\eqnref{equ:MIendm}, \eqnref{equ:CDreceivermED2} and \eqnref{equ:PDreceiverchmED2} can be simplified as
\begin{equation}
\label{equ:CDreceivermED3}
{{{\tilde{R}}}_{1}} ={{\tilde{X}}}+\sum_{k=1}^{K} B_k{\tilde{W}_k}+C{{\tilde{Z}}},
\end{equation}
\begin{align}
\label{equ:PDreceiverchmED3}
{{{R}}_{2}} \approx A\big|{{\tilde{X}}}\big|+\sum_{k=1}^{K} B^{'}_k{{W}_{||k}}+ C^{'}{N}.
\end{align}
From \eqnref{equ:CDreceivermED3} and \eqnref{equ:PDreceiverchmED3}, we can verify that  the real and imaginary parts of $C{{\tilde{Z}}}$ and $C^{'}{N}$ are  i.i.d. zero-mean real Gaussian random variables, which follow the same distribution $\mathcal{N}\big(0,\frac{C^2\sigma_{\textrm{cov}}^2}{2}\big)$.

To derive the MI, we  firstly define the following random variables as
\begin{equation}
\label{equ:Variables1}
{{{\tilde{X}}}_1} ={{\tilde{X}}},
\end{equation}
\begin{equation}
\label{equ:Variables2} {{{X}}_2} = A|{{\tilde{X}}}|.
\end{equation}
Due to the Markov chain ${{\tilde{X}}}\rightarrow ({{{\tilde{X}}}_1}, {{{X}}_2})\rightarrow ({{{\tilde{R}}}_1}, {{{R}}_2})$ and the smooth and uniquely invertible map from ${{\tilde{X}}}$ to  $({{{\tilde{X}}}_1}, {{{X}}_2})$, we have
\begin{equation}
\label{equ:MI}
\mathcal{I}({{\tilde{X}}};{{{\tilde{R}}}_1}, {{{R}}_2})=\mathcal{I}({{{\tilde{X}}}_1}, {{{X}}_2};{{{\tilde{R}}}_1}, {{{R}}_2}).
\end{equation}

We define a three-dimensional cone-normal (CN) coordinate system  $(c_I, c_Q, c_M)$~\cite{WangLZ22} to calculate $\mathcal{I}({{{\tilde{X}}}_1}, {{{X}}_2};{{{\tilde{R}}}_1}, {{{R}}_2})$, where $c_I$, $c_Q$, and $c_M$ denote the three axes of Cartesian coordinate system of the in-phase-quadrature-magnitude (I-Q-M) space, and
\begin{equation}
\label{equ:PN}
c_M=A\sqrt{c_I^2+c_Q^2}.
\end{equation}
In the CN coordinate system, the point $(c_1,c_2, c_3)$ is further expressed as $(\tilde{a},l)$, where $\tilde{a}$  and $|l|$ denote the nearest point on  the cone $\mathcal{U}$ to $(c_1,c_2, c_3)$ and the distance, respectively.
We can prove that the points $({{{\tilde{X}}}_1}, {{{X}}_2})$ lies on $\mathcal{U}$.
The random variables $({{{\tilde{X}}}_1}, {{{X}}_2})$ and $({{{\tilde{R}}}_1}, {{{R}}_2})$ in the Cartesian coordinate system is represented as $(\tilde{A}_{{{\tilde{X}}}},L_{\tilde{X}})$ and $(\tilde{A}_{{{\tilde{X}}},\tilde{W}, \tilde{Z}, N},L_{\tilde{X},\tilde{W}, \tilde{Z}, N})$ in the CN coordinate system, respectively. Thus, the MI is rewritten as
\begin{equation}
\label{equ:MIPN}
\mathcal{I}({{{\tilde{X}}}_1}, \!{{{X}}_2};\!{{{\tilde{R}}}_1}, \!{{{R}}_2})\!\!=\!\!\mathcal{I}(\tilde{A}_{{{\tilde{X}}}},L_{\tilde{X}};\tilde{A}_{{{\tilde{X}}},\tilde{W}, \tilde{Z}, N},L_{\tilde{X},\tilde{W}, \tilde{Z}, N}),
\end{equation}
Following the similar steps in~\cite{WangLZ22}, the asymptotic MI in \eqnref{equ:MIPN} can be simplified as
\begin{align}
\label{equ:MIPN2ED}
&\mathcal{I}({{{\tilde{X}}}_1}, {{{X}}_2};{{{\tilde{R}}}_1}, {{{R}}_2}) =  \mathcal{H}(\tilde{A}_{{{\tilde{X}}},\tilde{W}})-\mathcal{H}(\tilde{A}_{{{\tilde{X}}},\tilde{W}, \tilde{Z}, N}|\tilde{A}_{{{\tilde{X}}}}).
\end{align}

Then,  $\mathcal{H}(\tilde{A}_{{{\tilde{X}}},\tilde{W}})$ and $\mathcal{H}(\tilde{A}_{{{\tilde{X}}},\tilde{W}, \tilde{Z}, N}|\tilde{A}_{{{\tilde{X}}}})$ are calculated as follows.

1) $\mathcal{H}(\tilde{A}_{{{\tilde{X}}},\tilde{W}})$:
Since $\tilde{X}$ is zero-mean complex Gaussian random variables, the analysis of $\mathcal{H}(\tilde{A}_{{{\tilde{X}}},\tilde{W}})$ is similar to~\cite{WangLZ22}, which is calculated as
\begin{align}
\label{equ:MIAxwED}
&\mathcal{H}(\tilde{A}_{{{\tilde{X}}},\tilde{W}})=\frac{1}{\ln2}+\log_2\left(\pi\sqrt{A^2+1}\right).
\end{align}

2) Asymptotic $\mathcal{H}(\tilde{A}_{{{\tilde{X}}},\tilde{W}, \tilde{Z}, N}|\tilde{A}_{{{\tilde{X}}}})$: The conditional entropy can be written as
\begin{align}
\label{equ:MIAxwzn}
\mathcal{H}\big(\tilde{A}_{{{\tilde{X}}},\tilde{W}, \tilde{Z}, N}|\tilde{A}_{{{\tilde{X}}}}\big)&=\mathbb{E}_{\tilde{A}_{{\tilde{X}}}}[\mathcal{H}(\tilde{A}_{{{\tilde{X}}},\tilde{W}, \tilde{Z}, N}|\tilde{A}_{{{\tilde{X}}}}=\tilde{a}_{\tilde{X}}]\notag\\
&=\mathbb{E}_{\tilde{X}}[\mathcal{H}(\tilde{A}_{{{\tilde{X}}},\tilde{W}, \tilde{Z}, N}|{{{\tilde{X}}}}].
\end{align}
Given $\tilde{X}$, the entropy $\mathcal{H}\big(\tilde{A}_{{{\tilde{X}}},\tilde{W}, \tilde{Z}, N}|\tilde{A}_{{{\tilde{X}}}}\big)$ is independent with the phase of the complex Gaussian random variable $\tilde{X}$. Thus, without loss of generality, we define $\tilde{X}$ as $(X_I,0)$, where $X_I\geq 0$.
Then, we introduce ${{{\tilde{X}}}'_1}$ and ${{{X}}'_2}$ as
\begin{equation}
\label{equ:Variables3CN}
{{{\tilde{X}}}'_1} = {{\tilde{X}}}+\sum_{k=1}^{K} B_k{\tilde{W}_k},
\end{equation}
\begin{equation}
\label{equ:Variables4CN}
{{{X}}'_2} = A\big|{{\tilde{X}}}\big|+\sum_{k=1}^{K} B^{'}_k{{W}_{||k}}.
\end{equation}
From \eqnref{equ:Variables3CN} and \eqnref{equ:Variables4CN}, the point $({{{\tilde{X}}}'_1},{{X}'_2})$ lies on the cone $\mathcal{U}$ in the I-Q-M space, which is approximated as
\begin{align}
\label{equ:Variables8CN}
({{{\tilde{X}}}'_1},{{X}'_2})&\approx ({{{\tilde{X}}}_1},{{X}_2})\!+\!\biggl(\sum_{k=1}^{K} B_k{\tilde{W}_k},\sum_{k=1}^{K} B^{'}_k{{W}_{||k}}\biggr)\notag\\
&=({{{\tilde{X}}}_1},{{X}_2})\!+\!\sqrt{\sum_{k=1}^{K} \bigg(\frac{B_k}{\sqrt{1+A^2}}+\frac{B'_kA}{\sqrt{1+A^2}} \bigg)^2}{{W_{||}}}\textbf{i}_{IM}+\sqrt{\sum_{k=1}^{K} B^2_k}{W_{\perp}}\textbf{i}_Q,
\end{align}
where ${{W}_{\perp}}$ and ${{W}_{||}}$  are the projections of ${\tilde{W}}$  onto the same direction and the vertical direction of ${{\tilde{X}}}$, respectively. ${{W}_{\perp}}$ and ${{W}_{||}}$ follow the same distributions as  ${{W}_{\perp k}}$ and ${{W}_{||k}}$, respectively. $\textbf{i}_{IM}$ and $\textbf{i}_{Q}$ are mutually orthogonal unit vectors defined as
\begin{align}
\label{equ:unvec}
&\textbf{i}_{IM}\triangleq \biggl(\frac{A}{\sqrt{1+A^2}},0,\frac{1}{\sqrt{1+A^2}}\biggr),\notag\\
&\textbf{i}_{Q}\triangleq (0,1,0).
\end{align}

As $P$ increases, the approximated point $({{{\tilde{X}}}'_1},{{X}'_2})$ in \eqref{equ:Variables8CN} lies on the tangent plane of $\mathcal{U}$ of the point $({{{\tilde{X}}}_1},{{X}_2})$, $\forall W_{||},W_{\perp}\in \mathbb{R}$.
Defining $\big({C\tilde{Z}},{C' N}\big)\triangleq{Z_1}\textbf{i}_{IM}+{Z_2}\textbf{i}_{Q}+{Z_3}\textbf{i}_{IQM}$, where $Z_1$, $Z_2$, and $Z_3$ are independent with each other and meet the same distributions $\mathcal{N}\big(0,\frac{C^2\sigma_{\textrm{cov}}^2}{2}\big)$.
Then, taking \eqnref{equ:Variables8CN} into \eqnref{equ:CDreceivermED3} and \eqnref{equ:PDreceiverchmED3}, we obtain
\begin{align}
\label{equ:sprechED}
(\tilde{R}_1,R_2)&=({{{\tilde{X}}}'_1},{{X}'_2})+\biggl({C\tilde{Z}},{C' N}\biggr)\notag\\
&\approx ({{{\tilde{X}}}_1},{{X}_2})\!\!+\!\!\left(\sqrt{{\sum_{k=1}^{K} \bigg(\!\frac{B_k}{\sqrt{1+A^2}}\!\!+\!\!\frac{B'_kA}{\sqrt{1+A^2}} \!\bigg)^2}}{{W_{||}}}\!+\!{Z_1} \right)\!\textbf{i}_{IM}\notag\\
&\,\,\,\,\,\,\,+\left( \sqrt{\sum_{k=1}^{K} B^2_k}{{W_{\perp}}}\!\!+\!\!{ Z_2}\right)\textbf{i}_Q+{Z_3}\textbf{i}_{IQM},
\end{align}
where $\textbf{i}_{IQM}$ denotes a unit vector, which is orthogonal to $\textbf{i}_{IM}$ and $\textbf{i}_Q$.

In the high SNR regime, $(\tilde{R}_1,R_2)$ converges to $({{{\tilde{X}}}_1},{{X}_2})$ on $\mathcal{U}$ in probability. As a result, $\tilde{A}_{{{\tilde{X}}},\tilde{W}, \tilde{Z}, N}$ converges to the projection on the talent plane of the point $({{{\tilde{X}}}_1},{{X}_2})$ on $\mathcal{U}$, called as $\mathcal{S}$. Based on \eqnref{equ:sprechED}, $\tilde{A}_{{{\tilde{X}}},\tilde{W}, \tilde{Z}, N}$ is approximated as
\begin{align}
\label{equ:AxwznappED}
\tilde{A}_{{{\tilde{X}}},\tilde{W}, \tilde{Z}, N}&\approx({{{\tilde{X}}}_1},{{X}_2})+\left( \sqrt{\sum_{k=1}^{K} B^2_k}{{W_{\perp}}}+{Z_2}\right)\textbf{i}_Q \notag\\
& +\left(\sqrt{{\sum_{k=1}^{K} \bigg(\frac{B_k}{\sqrt{1+A^2}}+\frac{B'_kA}{\sqrt{1+A^2}} \bigg)^2}}{{W_{||}}}+{ Z_1} \right)\textbf{i}_{IM}.
\end{align}
Due to the fact that $W_{||}$, $W_{\perp}$, $Z_1$ and $Z_2$ are i.i.d. Gaussian, the approximated $\tilde{A}_{{{\tilde{X}}},\tilde{W}, \tilde{Z}, N}$  is complex Gaussian on $\mathcal{S}$ with known $\tilde{X}$. The covariance matrix of $\tilde{A}_{{{\tilde{X}}},\tilde{W}, \tilde{Z}, N}$ is given by
\begin{align}
\label{equ:VarED}
\mathcal{G}=\!\!\left[\begin{array}{cccc}
    \frac{{\sum_{k=1}^{K} \bigg(\frac{B_k}{\sqrt{1+A^2}}+\frac{B'_kA}{\sqrt{1+A^2}} \bigg)^2}\sigma_{\textrm{A}}^2+C^2\sigma_{\textrm{cov}}^2}{2} & \!\!\! 0 \!\!\!\\
    \!\!\!0 \!\!\! & \!\!\!\!\!   \frac{\sum_{k=1}^{K} B^2_k\sigma_{\textrm{A}}^2+C^2\sigma_{\textrm{cov}}^2}{2}\\
\end{array}\right].
\end{align}
From \eqnref{equ:VarED}, the approximated conditional entropy $\mathcal{H}(\tilde{A}_{{{\tilde{X}}},\tilde{W}, \tilde{Z}, N}|\tilde{A}_{{{\tilde{X}}}})$ is
\begin{align}
\label{equ:MIAxwznappED}
\mathcal{H}(\tilde{A}_{{{\tilde{X}}},\tilde{W}, \tilde{Z}, N}|\tilde{A}_{{{\tilde{X}}}})&=\mathbb{E}_{\tilde{X}}[\mathcal{H}(\tilde{A}_{{{\tilde{X}}},\tilde{W}, \tilde{Z}, N}|\tilde{A}_{{{\tilde{X}}}}=\tilde{a}_{\tilde{X}}]\notag\\
&=  \log_2\pi e+\frac{1}{2}\log_2\left( {\sum_{k=1}^{K} B^2_k\sigma_{\textrm{A}}^2+C^2\sigma_{\textrm{cov}}^2}\right)\notag\\
&\,\,+\frac{1}{2}\log_2\left({{\sum_{k=1}^{K} \bigg(\frac{B_k}{\sqrt{1+A^2}}+\frac{B'_kA}{\sqrt{1+A^2}} \bigg)^2}\sigma_{\textrm{A}}^2+C^2\sigma_{\textrm{cov}}^2} \right).
\end{align}

3) Asymptotic $\mathcal{I}({{\tilde{X}}};{{{\tilde{R}}}_1}, {{{R}}_2})$:
Taking \eqnref{equ:MIAxwED} and \eqnref{equ:MIAxwznappED} into \eqnref{equ:MIPN2ED}, the asymptotic MI in~\eqnref{equ:MIendm} is obtained.

\section{ Proof of Proposition \ref{proposition:prop4}}
Taking  $A$, $B_k$, $C$, $A^{'}$, $B^{'}_k$, $C^{'}$ and $\gamma$ defined in Proposition  \ref{proposition:prop3} into \eqnref{equ:MIendm}, the approximated MI is rewritten as
\begin{align}
\label{equ:MIendsi}
&\mathcal{I}({{{\tilde{X}}}};{{{\tilde{R}}}_1}, {{{R}}_2})\approx\frac{1}{2}\log_2\left({\frac{{{\sum_{k=1}^{K} \frac{\alpha^2_k}{\rho_k P |\tilde{h}_k|^2}}}(\sum_{k'=1}^{K}\beta_{k'})^2\sigma_{\textrm{cov}}^2}{2\sum_{k=1}^{K} \frac{\beta^2_k}{(1-\rho_k ) P |\tilde{h}_k|^2}(\sum_{k'=1}^{K}\alpha_{k'})^2\sigma_{\textrm{rec}}^2}+1}\right)\notag\\
&-\frac{1}{2}\log_2\bigg(\sum_{k=1}^{K}\bigg(\frac{\alpha_k}{\sqrt{P}|\tilde{h}_k|}\bigg)^2\frac{\sigma_{\textrm{A}}^2}{(\sum_{k'=1}^{K}\alpha_{k'})^2} +\sum_{k=1}^{K} \frac{\alpha^2_k}{\rho_k P |\tilde{h}_k|^2}\frac{\sigma_{\textrm{cov}}^2}{(\sum_{k'=1}^{K}\alpha_{k'})^2}\bigg)\notag\\
&-\frac{1}{2}\log_2\bigg(\frac{1}{{{\frac{{{\sum_{k=1}^{K} \frac{\alpha^2_k}{\rho_k P |\tilde{h}_k|^2}}}(\sum_{k'=1}^{K}\beta_{k'})^2\sigma_{\textrm{cov}}^2}{2\sum_{k=1}^{K} \frac{\beta^2_k}{(1-\rho_k ) P |\tilde{h}_k|^2}(\sum_{k'=1}^{K}\alpha_{k'})^2\sigma_{\textrm{rec}}^2}+1}}}
\sum_{k=1}^{K}\bigg(\frac{\alpha_k}{(\sum_{k'=1}^{K} \alpha_{k'})\sqrt{P}|\tilde{h}_k|}\notag\\
&+\frac{\beta_k}{\sqrt{P}|\tilde{h}_k|}\frac{{{\sum_{k''=1}^{K} \frac{\alpha^2_{k''}}{\rho_{k''} P |\tilde{h}_{k''}|^2}}}(\sum_{k'=1}^{K}\beta_{k'})\sigma_{\textrm{cov}}^2}{2\sum_{k''=1}^{K} \frac{\beta^2_{k''}}{(1-\rho_{k''} ) P |\tilde{h}_{k''}|^2}(\sum_{k'=1}^{K}\alpha_{k'})^2\sigma_{\textrm{rec}}^2}\bigg)^2\sigma_{\textrm{A}}^2
+\sum_{k=1}^{K} \frac{\alpha^2_k}{\rho_k P |\tilde{h}_k|^2}\frac{\sigma_{\textrm{cov}}^2}{(\sum_{k'=1}^{K}\alpha_{k'})^2} \bigg).
\end{align}
Let $\mu=\sum_{k=1}^{K}\frac{ \alpha^2_k}{\rho_k|\tilde{h}_k|^2}$, $\kappa=\sum_{k=1}^{K}\frac{\beta^2_k}{(1-\rho_k)|\tilde{h}_k|^2}$, $\varepsilon=\sum_{k=1}^{K}\alpha_{k}$, $\zeta=\sum_{k=1}^{K}\beta_{k}$, $\xi=\sum_{k=1}^{K}\frac{2\alpha_k\beta_k}{|\tilde{h}_k|^2}$, and $\varpi=\sum_{k=1}^{K}\frac{ \alpha^2_k}{|\tilde{h}_k|^2}$.
Then, we have  $\mathcal{I}({{{\tilde{X}}}};{{{\tilde{R}}}_1}, {{{R}}_2})=f(\boldsymbol{\rho},\boldsymbol{\alpha},\boldsymbol{\beta}) =f(\mu,\kappa,\varepsilon,\zeta,\xi,\varpi)$.
To obtain the maximum MI value, we need to solve the extreme point of the function $f(\mu,\kappa,\varepsilon,\zeta,\xi,\varpi)$.  The derivatives of the function $f(\mu,\kappa,\varepsilon,\zeta,\xi,\varpi)$ with respect to $\rho_k$, $\alpha_k$ and $\beta_k$ are respectively given by
\begin{align}
\label{equ:MIendsim32}
&\frac{\partial f(\mu,\kappa,\varepsilon,\zeta,\xi,\varpi)}{\partial \rho_k}=\frac{\partial f(\mu,\kappa,\varepsilon,\zeta,\xi,\varpi)}{\partial\mu}\frac{\partial \mu}{\partial\rho_k} +\frac{\partial f(\mu,\kappa,\varepsilon,\zeta,\xi,\varpi)}{\partial\kappa}\frac{\partial \kappa}{\partial\rho_k}\notag\\
&=-\frac{\partial f(\mu,\kappa,\varepsilon,\zeta,\xi,\varpi)}{\partial\mu}\frac{\alpha_k^2}{\rho_k^2|\tilde{h}_k|^2}
+\frac{\partial f(\mu,\kappa,\varepsilon,\zeta,\xi,\varpi)}{\partial\kappa}\frac{\beta_k^2}{(1-\rho_k^2)|\tilde{h}_k|^2},
\end{align}
\begin{align}
\label{equ:MIendsim33}
\frac{\partial f(\mu,\kappa,\varepsilon,\zeta,\xi,\varpi)}{\partial \alpha_k}&=
\frac{\partial f(\mu,\kappa,\varepsilon,\zeta,\xi,\varpi)}{\partial\mu}\frac{\partial \mu}{\partial\alpha_k}+\frac{\partial f(\mu,\kappa,\varepsilon,\zeta,\xi,\varpi)}{\partial\xi}\frac{\partial \xi}{\partial\alpha_k}\notag\\
&+\frac{\partial f(\mu,\kappa,\varepsilon,\zeta,\xi,\varpi)}{\partial\varepsilon}\frac{\partial \varepsilon}{\partial\alpha_k}+\frac{\partial f(\mu,\kappa,\varepsilon,\zeta,\xi,\varpi)}{\partial\varpi}\frac{\partial \varpi}{\partial\alpha_k}\notag\\
&=\frac{\partial f(\mu,\kappa,\varepsilon,\zeta,\xi,\varpi)}{\partial\mu}\frac{2\alpha_k}{\rho_k |\tilde{h}_k|^2}+\frac{\partial f(\mu,\kappa,\varepsilon,\zeta,\xi,\varpi)}{\partial\xi}\frac{2\beta_k}{|\tilde{h}_k|^2}\notag\\
&+\frac{\partial f(\mu,\kappa,\varepsilon,\zeta,\xi,\varpi)}{\partial\varepsilon}+\frac{\partial f(\mu,\kappa,\varepsilon,\zeta,\xi,\varpi)}{\partial\varpi}\frac{2\alpha_k}{|\tilde{h}_k|^2},
\end{align}
and
\begin{align}
\label{equ:MIendsim34}
&\frac{\partial f(\mu,\kappa,\varepsilon,\zeta,\xi,\varpi)}{\partial \beta_k}=\frac{\partial f(\mu,\kappa,\varepsilon,\zeta,\xi,\varpi)}{\partial\kappa}\frac{\partial \kappa}{\partial\beta_k}+\frac{\partial f(\mu,\kappa,\varepsilon,\zeta,\xi,\varpi)}{\partial\xi}\frac{\partial \xi}{\partial\beta_k}+\frac{\partial f(\mu,\kappa,\varepsilon,\zeta,\xi,\varpi)}{\partial\zeta}\frac{\partial \zeta}{\partial\beta_k}\notag\\
&=\frac{\partial f(\mu,\kappa,\varepsilon,\zeta,\xi,\varpi)}{\partial\kappa}\frac{2\beta_k}{(1-\rho_k)|\tilde{h}_k|^2}+
\frac{\partial f(\mu,\kappa,\varepsilon,\zeta,\xi,\varpi)}{\partial\xi}\frac{2\alpha_k}{|\tilde{h}_k|^2}+\frac{\partial f(\mu,\kappa,\varepsilon,\zeta,\xi,\varpi)}{\partial\zeta}.
\end{align}

From \eqnref{equ:MIendsim32}, when $\frac{\partial f(\mu,\kappa,\varepsilon,\zeta,\xi,\varpi)}{\partial\rho_k}=0$, we have
\begin{align}
\label{equ:MIendsim35}
&\beta_k=\sqrt{\frac{\frac{\partial f(\mu,\kappa,\varepsilon,\zeta,\xi,\varpi)}{\partial\mu}}{\frac{\partial f(\mu,\kappa,\varepsilon,\zeta,\xi,\varpi)}{\partial\kappa}}}\frac{(1-\rho_k)\alpha_k}{\rho_k}.
\end{align}
From~\eqnref{equ:MIendsim34}, when $\frac{\partial f(\mu,\kappa,\varepsilon,\zeta,\xi)}{\partial\beta_k}=0$, we have
\begin{align}
\label{equ:MIendsim36}
&\frac{\partial f(\mu,\kappa,\varepsilon,\zeta,\xi,\varpi)}{\partial\varpi}\frac{2\alpha_k}{|\tilde{h}_k|^2}=
-\frac{\partial f(\mu,\kappa,\varepsilon,\zeta,\xi,\varpi)}{\partial\kappa}\frac{2\beta_k}{(1-\rho_k)|\tilde{h}_k|^2}-\frac{\partial f(\mu,\kappa,\varepsilon,\zeta,\xi,\varpi)}{\partial\zeta}.
\end{align}
Taking \eqnref{equ:MIendsim35} into \eqnref{equ:MIendsim36}, we can prove that
\begin{align}
\label{equ:MIendsim37}
\frac{2\alpha_k}{|\tilde{h}_k|^2}=-\frac{\frac{\partial f(\mu,\kappa,\varepsilon,\zeta,\xi,\varpi)}{\partial\zeta}}{\bigg(\frac{\partial f(\mu,\kappa,\varepsilon,\zeta,\xi,\varpi)}{\partial\varpi}+\sqrt{\frac{\partial f(\mu,\kappa,\varepsilon,\zeta,\xi,\varpi)}{\partial\kappa}}\frac{\sqrt{{\frac{\partial f(\mu,\kappa,\varepsilon,\zeta,\xi,\varpi)}{\partial\mu}}}}{\rho_k}\bigg)}.
\end{align}
Taking \eqnref{equ:MIendsim37} into \eqnref{equ:MIendsim35}, we have
\begin{align}
\label{equ:MIendsim37-1}
\frac{2\beta_k}{|\tilde{h}_k|^2}=
-\sqrt{\frac{\frac{\partial f(\mu,\kappa,\varepsilon,\zeta,\xi,\varpi)}{\partial\mu}}{\frac{\partial f(\mu,\kappa,\varepsilon,\zeta,\xi,\varpi)}{\partial\kappa}}}\frac{(1-\rho_k)}{\rho_k}
&\frac{\frac{\partial f(\mu,\kappa,\varepsilon,\zeta,\xi,\varpi)}{\partial\zeta}}{\bigg(\frac{\partial f(\mu,\kappa,\varepsilon,\zeta,\xi,\varpi)}{\partial\varpi}+\sqrt{\frac{\partial f(\mu,\kappa,\varepsilon,\zeta,\xi,\varpi)}{\partial\kappa}}\frac{\sqrt{{\frac{\partial f(\mu,\kappa,\varepsilon,\zeta,\xi,\varpi)}{\partial\mu}}}}{\rho_k}\bigg)}.
\end{align}

Taking \eqnref{equ:MIendsim37} and \eqnref{equ:MIendsim37-1} into \eqnref{equ:MIendsim33}, we can obtain $\rho_k$.  Since the derivative of the function $f(\mu,\kappa,\varepsilon,\zeta,\xi,\varpi)$ is independent of $k$, the splitting ratios $[\rho_1,\rho_2,\cdots, \rho_K]$ at different antennas are identical.
When $\rho_1=\rho_2=\cdots=\rho_K=\rho$, the approximated MI in \eqnref{equ:MIendsi} is simplified as
\begin{align}
\label{equ:MIendsim25cc}
&\mathcal{I}({{{\tilde{X}}}};{{{\tilde{R}}}_1}, {{{R}}_2})\approx\frac{1}{2}\log_2\left({\frac{{{\sum_{k=1}^{K} \frac{\alpha^2_k}{\rho P |\tilde{h}_k|^2}}}(\sum_{k'=1}^{K}\beta_{k'})^2\sigma_{\textrm{cov}}^2}{2\sum_{k=1}^{K} \frac{\beta^2_k}{(1-\rho ) P |\tilde{h}_k|^2}(\sum_{k'=1}^{K}\alpha_{k'})^2\sigma_{\textrm{rec}}^2}+1}\right)\notag\\
&-\frac{1}{2}\log_2\bigg(\sum_{k=1}^{K}\bigg(\frac{\alpha_k}{\sqrt{P}|\tilde{h}_k|}\bigg)^2\frac{\sigma_{\textrm{A}}^2}{(\sum_{k'=1}^{K}\alpha_{k'})^2} +\sum_{k=1}^{K} \frac{\alpha^2_k}{\rho P |\tilde{h}_k|^2}\frac{\sigma_{\textrm{cov}}^2}{(\sum_{k'=1}^{K}\alpha_{k'})^2}\bigg)\notag\\
&-\frac{1}{2}\log_2\bigg(\frac{1}{{{\frac{{{\sum_{k=1}^{K} \frac{\alpha^2_k}{\rho P |\tilde{h}_k|^2}}}(\sum_{k'=1}^{K}\beta_{k'})^2\sigma_{\textrm{cov}}^2}{2\sum_{k=1}^{K} \frac{\beta^2_k}{(1-\rho ) P |\tilde{h}_k|^2}(\sum_{k'=1}^{K}\alpha_{k'})^2\sigma_{\textrm{rec}}^2}+1}}}\sum_{k=1}^{K}\bigg(\frac{\alpha_k}{(\sum_{k'=1}^{K} \alpha_{k'})\sqrt{P}|\tilde{h}_k|}\notag\\
&+\frac{\beta_k}{\sqrt{P}|\tilde{h}_k|}\frac{{{\sum_{k''=1}^{K} \frac{\alpha^2_{k''}}{\rho P |\tilde{h}_{k''}|^2}}}(\sum_{k'=1}^{K}\beta_{k'})\sigma_{\textrm{cov}}^2}{2\sum_{k''=1}^{K} \frac{\beta^2_{k''}}{(1-\rho ) P |\tilde{h}_{k''}|^2}(\sum_{k'=1}^{K}\alpha_{k'})^2\sigma_{\textrm{rec}}^2}\bigg)^2\sigma_{\textrm{A}}^2
+\sum_{k=1}^{K} \frac{\alpha^2_k}{\rho P |\tilde{h}_k|^2}\frac{\sigma_{\textrm{cov}}^2}{(\sum_{k'=1}^{K}\alpha_{k'})^2} \bigg).
\end{align}
Redefining $\mu=\sum_{k=1}^{K}\frac{\alpha^2_k}{|\tilde{h}_k|^2}$, $\kappa=\sum_{k=1}^{K}\frac{\beta^2_k}{|\tilde{h}_k|^2}$, $\varepsilon=\sum_{k=1}^{K}\alpha_{k}$, $\zeta=\sum_{k=1}^{K}\beta_{k}$, and $\xi=\sum_{k=1}^{K}\frac{2\alpha_k\beta_k}{|\tilde{h}_k|^2}$,
we have $\mathcal{I}({{{\tilde{X}}}};{{{\tilde{R}}}_1}, {{{R}}_2})=f(\rho, \boldsymbol{\alpha},\boldsymbol{\beta}) =f(\mu,\kappa,\varepsilon,\zeta,\xi)$.
The derivative of the function $f(\mu,\kappa,\varepsilon,\zeta,\xi)$ with respect to $\alpha_k$ is given by
\begin{align}
\label{equ:MIendsim26}
&\frac{\partial f(\mu,\kappa,\varepsilon,\zeta,\xi)}{\partial \alpha_k}=\frac{\partial f(\mu,\kappa,\varepsilon,\zeta,\xi)}{\partial\mu}\frac{\partial \mu}{\partial\alpha_k}+\frac{\partial f(\mu,\kappa,\varepsilon,\zeta,\xi)}{\partial\xi}\frac{\partial \xi}{\partial\alpha_k}+\frac{\partial f(\mu,\kappa,\varepsilon,\zeta,\xi)}{\partial\varepsilon}\frac{\partial \varepsilon}{\partial\alpha_k}\notag\\
&=\frac{\partial f(\mu,\kappa,\varepsilon,\zeta,\xi)}{\partial\mu}\frac{2\alpha_k}{|\tilde{h}_k|^2}+\frac{\partial f(\mu,\kappa,\varepsilon,\zeta,\xi)}{\partial\xi}\frac{2\beta_k}{|\tilde{h}_k|^2}
+\frac{\partial f(\mu,\kappa,\varepsilon,\zeta,\xi)}{\partial\varepsilon}.
\end{align}
The derivative of the function $f(\mu,\kappa,\varepsilon,\zeta,\xi)$ with respect to  $\beta_k$ is given by
\begin{align}
\label{equ:MIendsim26-1}
&\frac{\partial f(\mu,\kappa,\varepsilon,\zeta,\xi)}{\partial \beta_k}=\frac{\partial f(\mu,\kappa,\varepsilon,\zeta,\xi)}{\partial\kappa}\frac{\partial \kappa}{\partial\beta_k}+\frac{\partial f(\mu,\kappa,\varepsilon,\zeta,\xi)}{\partial\xi}\frac{\partial \xi}{\partial\beta_k}+\frac{\partial f(\mu,\kappa,\varepsilon,\zeta,\xi)}{\partial\zeta}\frac{\partial \zeta}{\partial\beta_k}\notag\\
&=\frac{\partial f(\mu,\kappa,\varepsilon,\zeta,\xi)}{\partial\kappa}\frac{2\beta_k}{|\tilde{h}_k|^2}+\frac{\partial f(\mu,\kappa,\varepsilon,\zeta,\xi)}{\partial\xi}\frac{2\alpha_k}{|\tilde{h}_k|^2}
+\frac{\partial f(\mu,\kappa,\varepsilon,\zeta,\xi)}{\partial\zeta}.
\end{align}
By letting $\frac{\partial f(\mu,\kappa,\varepsilon,\zeta,\xi)}{\partial\alpha_k}=0$ and $\frac{\partial f(\mu,\kappa,\varepsilon,\zeta,\xi)}{\partial\beta_k}=0$, it directly follows that
\begin{align}
\label{equ:MIendsim27cc}
\frac{\alpha_k}{\beta_k}=\frac{\frac{\partial f(\mu,\kappa,\varepsilon,\zeta,\xi)}{\partial\xi}\frac{\partial f(\mu,\kappa,\varepsilon,\zeta,\xi)}{\partial\zeta}-\frac{\partial f(\mu,\kappa,\varepsilon,\zeta,\xi)}{\partial\varepsilon}\frac{\partial f(\mu,\kappa,\varepsilon,\zeta,\xi)}{\partial\kappa}}{\frac{\partial f(\mu,\kappa,\varepsilon,\zeta,\xi)}{\partial\xi}\frac{\partial f(\mu,\kappa,\varepsilon,\zeta,\xi)}{\partial\varepsilon}-\frac{\partial f(\mu,\kappa,\varepsilon,\zeta,\xi)}{\partial\mu}\frac{\partial f(\mu,\kappa,\varepsilon,\zeta,\xi)}{\partial\zeta}}.
\end{align}
From \eqnref{equ:MIendsim27cc}, we can see that  $\frac{\alpha_k}{\beta_k}$ is fixed and independent of $k$. Defining $\frac{\alpha_1}{\beta_1}=\frac{\alpha_2}{\beta_2}=\cdots=\frac{\alpha_K}{\beta_K}=\delta$, the approximated MI in \eqnref{equ:MIendsim25cc} is further simplified as
\begin{align}
\label{equ:MIendsi38}
\mathcal{I}({{{\tilde{X}}}};{{{\tilde{R}}}_1}, {{{R}}_2})
&\approx\log_2(P)-\frac{1}{2}\log_2\bigg(\bigg({{\frac{\sigma_{\textrm{cov}}^2}{\rho  }}}+\sigma_{\textrm{A}}^2 \bigg)
\bigg({\sigma_{\textrm{A}}^2} +\frac{{{\frac{\sigma_{\textrm{cov}}^2}{\rho}}}2 \frac{\sigma_{\textrm{rec}}^2}{(1-\rho ) }}{{{\frac{\sigma_{\textrm{cov}}^2}{\rho}}}+2 \frac{\sigma_{\textrm{rec}}^2}{(1-\rho ) }}\bigg)\bigg)\notag\\
&-\log_2\bigg(\sum_{k=1}^{K}\frac{\beta_k^2}{|\tilde{h}_k|^2}\bigg)+2\log_2\bigg(\sum_{k'=1}^{K}\beta_{k'}\bigg).
\end{align}
Let
\begin{align}
\label{equ:gbeta}
&g(\beta_1,\beta_2, \cdots, \beta_K)=-\log_2\bigg(\sum_{k=1}^{K}\frac{\beta_k^2}{|\tilde{h}_k|^2}\bigg)+2\log_2\bigg(\sum_{k'=1}^{K}\beta_{k'}\bigg),
\end{align}
and
\begin{align}
\label{equ:MIendsim20-11}
s(\rho)&=\bigg({{\frac{\sigma_{\textrm{cov}}^2}{\rho  }}}+\sigma_{\textrm{A}}^2 \bigg)
\bigg({\sigma_{\textrm{A}}^2} +\frac{{{\frac{\sigma_{\textrm{cov}}^2}{\rho}}}2 \frac{\sigma_{\textrm{rec}}^2}{(1-\rho ) }}{{{\frac{\sigma_{\textrm{cov}}^2}{\rho}}}+2 \frac{\sigma_{\textrm{rec}}^2}{(1-\rho ) }}\bigg)\notag\\
&=\frac{ (\rho\sigma_{\textrm{A}}^2 + \sigma_{\textrm{cov}}^2) ((-1 + \rho) \sigma_{\textrm{A}}^2 \sigma_{\textrm{cov}}^2 - 2 \rho \sigma_{\textrm{A}}^2 \sigma_{\textrm{rec}}^2 -
   2 \sigma_{\textrm{cov}}^2 \sigma_{\textrm{rec}}^2)}{\rho ((-1 + \rho) \sigma_{\textrm{cov}}^2 - 2 \rho \sigma_{\textrm{rec}}^2)}.
\end{align}
The approximated MI in \eqnref{equ:MIendsi38} is rewritten as
\begin{align}
\label{equ:MIendsi38-1}
&\mathcal{I}({{{\tilde{X}}}};{{{\tilde{R}}}_1}, {{{R}}_2})\approx\log_2(P)-\frac{1}{2}\log_2\big(s(\rho)\big)
+g( \beta_1,\beta_2, \cdots,\beta_k).
\end{align}
From \eqnref{equ:MIendsi38-1}, the derivative of the function $s(\rho)$ is given by
\begin{align}
\label{equ:MIendsim21}
\frac{\textrm{ d} s(\rho)}{\textrm{d} \rho}&=-\frac{\sigma_{\textrm{cov}}^2 (2 \sigma_{\textrm{cov}}^2 \sigma_{\textrm{rec}}^2 (\sigma_{\textrm{cov}}^2 \!\!- \!\!2 \rho \sigma_{\textrm{cov}}^2\!\! +\!\! 4 \rho \sigma_{\textrm{rec}}^2)}{\rho^2 ((-1 + \rho) \sigma_{\textrm{cov}}^2 - 2 \rho \sigma_{\textrm{rec}}^2)^2}\notag\\
&-\frac{\sigma_{\textrm{A}}^2((-1 \!\!+ \!\!\rho)^2 \sigma_{\textrm{cov}}^4\!\! +\!\! 2 (2 \!\!-\!\! 3 \rho) \rho \sigma_{\textrm{cov}}^2 \sigma_{\textrm{rec}}^2\!\! +\!\!
       8 \rho^2 \sigma_{\textrm{rec}}^4))}{\rho^2 ((-1 + \rho) \sigma_{\textrm{cov}}^2 - 2 \rho \sigma_{\textrm{rec}}^2)^2}.
\end{align}
The function $\frac{\textrm{ d} s(\rho)}{\textrm{d} \rho}=0$ has two roots $\Upsilon$ and
\begin{align}
\label{equ:MIendsim23-11}
\Phi=\frac{\sigma_{\textrm{cov}}^2 (\sigma_{\textrm{cov}}^2 - 2 \sigma_{\textrm{rec}}^2) (\sigma_{\textrm{A}}^2 + 2\sigma_{\textrm{rec}}^2)+
  \sqrt{2
  \Psi}}{\sigma_{\textrm{A}}^2 (\sigma_{\textrm{cov}}^2 - 4 \sigma_{\textrm{rec}}^2) (\sigma_{\textrm{cov}}^2 - 2 \sigma_{\textrm{rec}}^2)},
\end{align}
where $\Upsilon$ and $\Psi$ are defined in Proposition  \ref{proposition:prop4}.
We can prove that $\Upsilon$ is the root of interest that meets $0<\Upsilon<1$ when $\sigma_{\textrm{cov}}^2 >4 \sigma_{\textrm{rec}}^2$.

The second derivative of the  function $s(\rho)$ with respect $\rho$ is expressed as
\begin{align}
\label{equ:MIendsim24}
\frac{\textrm{d}^2 s(\rho)}{\textrm{d}\rho^2}|_{\rho=\Upsilon}&=
\bigg(16 \sigma_{\textrm{A}}^8 \sigma_{\textrm{cov}}^6(\sigma_{\textrm{cov}}^2\!\! -\!\! 4 \sigma_{\textrm{rec}}^2)^4 (\sigma_{\textrm{cov}}^2 \!\!- \!\!2 \sigma_{\textrm{rec}}^2)^3 (\sigma_{\textrm{A}}^2 \!\!+\!\! \sigma_{\textrm{cov}}^2 \!\!-\!\! 2 \sigma_{\textrm{rec}}^2) \sigma_{\textrm{rec}}^2\bigg) \notag\\
&\frac{-(\sigma_{\textrm{A}}^2 +
       2 \sigma_{\textrm{rec}}^2) (\sigma_{\textrm{A}}^2 \sigma_{\textrm{cov}}^4 + 4 \sigma_{\textrm{cov}}^4 \sigma_{\textrm{rec}}^2 - 8 \sigma_{\textrm{cov}}^2 \sigma_{\textrm{rec}}^4 -
       2 \sqrt{\Psi})}{(-\sigma_{\textrm{A}}^2 \sigma_{\textrm{cov}}^2 (\sigma_{\textrm{cov}}^2 - 2 \sigma_{\textrm{rec}}^2) +\Gamma)^3 (-2 \sigma_{\textrm{A}}^2 \sigma_{\textrm{cov}}^2 \sigma_{\textrm{rec}}^2 +\Gamma)^3},
\end{align}
where $\Gamma=- 2 \sigma_{\textrm{cov}}^4 \sigma_{\textrm{rec}}^2 + 4 \sigma_{\textrm{cov}}^2 \sigma_{\textrm{rec}}^4 +
       \sqrt{\Psi}$.
 We can prove that $\frac{\textrm{d}^2 s(\rho)}{\textrm{d}\rho^2}|_{\rho=\Upsilon}>0$ when $\sigma_{\textrm{cov}}^2 >4 \sigma_{\textrm{rec}}^2$. Defining $S(\rho)= -\frac{1}{2}\log_2(s(\rho))$, we have
 \begin{align}
\label{equ:MIendsim24-1}
 \frac{\textrm{d}^2 S(\rho)}{\textrm{d}\rho^2}|_{\rho=\Upsilon}&= -\frac{1}{2}\log_2(s(\rho))|_{\rho=\Upsilon}\notag\\
 &=-\frac{1}{2\ln2}\frac{\frac{\textrm{d}^2 s(\rho)}{\textrm{d}\rho^2}s(\rho)-(\frac{\textrm{d} s(\rho)}{\textrm{d}\rho})^2}{(s(\rho))^2}.
\end{align}
Since $\frac{\textrm{d}^2 s(\rho)}{\textrm{d}\rho^2}|_{\rho=\Upsilon}>0, \frac{\textrm{d} s(\rho)}{\textrm{d}\rho}|_{\rho=\Upsilon}=0$, and $s(\rho)|_{\rho=\Upsilon}>0$, we have $\frac{\textrm{d}^2 S(\rho)}{\textrm{d}\rho^2}|_{\rho=\Upsilon}<0$.

On the other hand, the derivative of the function $g(\beta_1,\beta_2, \cdots,\beta_K)$ with respect to $\beta_k$ is given by
\begin{align}
\label{equ:MIendsi30}
&\frac{\partial g( \beta_1,\beta_2, \cdots,\beta_K)}{\partial \beta_k}=
-\frac{\frac{2\beta_k}{|\tilde{h}_k|^2}}{(\sum_{k''=1}^{K}\frac{\beta_{k''}^2}{|\tilde{h}_{k''}|^2})\ln2}+\frac{2}{(\sum_{k'=1}^{K}\beta_{k'})\ln2}.
\end{align}
Letting $\frac{\partial g(\beta_1,\beta_2,\cdots, \beta_K)}{\partial\beta_k}=0$,  we can obtain that
 \begin{align}
\label{equ:MIendsi30-2}
\frac{\beta_k}{|\tilde{h}_k|^2}=\frac{\sum_{k''=1}^{K}\frac{\beta_{k''}^2}{|\tilde{h}_{k''}|^2}}{\sum_{k'=1}^{K}\beta_{k'}}, k={1,2,\cdots,K}.
\end{align}
Based on \eqnref{equ:MIendsi30-2}, we have
  \begin{align}
\label{equ:MIendsi33}
 \frac{\beta_k}{|\tilde{h}_k|^2}=\frac{\beta_j}{|\tilde{h}_j|^2},
 \end{align}
 for  arbitrary different $k$ and $j$. According to \eqnref{equ:MIendsim27cc} and \eqnref{equ:MIendsi33}, the combining coefficients $\alpha_k$ and $\beta_k$ in~\eqnref{equ:optCC1} are respectively achieved.

The second derivative of the function $g(\beta_1,\beta_2, \cdots, \beta_K)$ with respect to $\beta_k$ is given by
 \begin{align}
\label{equ:MIendsi31}
\frac{\partial^2 g( \beta_1,\beta_2, \cdots,\beta_K)}{\partial \beta^2_k}&=-\frac{\frac{2}{|\tilde{h}_k|^2}\sum_{k''=1}^{K}\frac{\beta_{k''}^2}{|\tilde{h}_{k''}|^2}-\frac{2\beta_k}{|\tilde{h}_k|^2}\frac{2\beta_k}{|\tilde{h}_k|^2}}{(\sum_{k''=1}^{K}\frac{\beta_{k''}^2}{|\tilde{h}_{k''}|^2})^2\ln2}
-\frac{2}{(\sum_{k'=1}^{K}\beta_{k'})^2\ln2}.
\end{align}
From \eqnref{equ:optCC1}, when $\beta_k=c_{\beta}|\tilde{h}_k|^2$ for all $k$, \eqnref{equ:MIendsi31} is simplified as
 \begin{align}
\label{equ:MIendsi31-1}
\frac{\partial^2 g( \beta_1,\beta_2, \cdots,\beta_K)}{\partial \beta^2_k}=\frac{-\frac{2}{|\tilde{h}_{k}|^2}\sum_{k'=1}^{K}|\tilde{h}_{k'}|^2+2}{c^2_{\beta}\sum_{k'=1}^{K}|\tilde{h}_{k'}|^2\ln2 }\leq0.
\end{align}

Therefore, the Hessian matrix of $\mathcal{I}({{{\tilde{X}}}};{{{\tilde{R}}}_1}, {{{R}}_2})$ in \eqnref{equ:MIendsi38-1} at point $(\rho, \beta_1,\beta_2,\cdots,\beta_K)$ is expressed as
\begin{align}
\label{equ:Hess}
&H(\rho, \beta_1,\beta_2, \cdots,\beta_K)=
     \left(
\begin{matrix}
 \frac{\partial^2 S( \rho)}{\partial \rho^2}     &0      & \cdots & 0     \\
 0      & \frac{\partial^2 g( \beta_1,\beta_2, \cdots,\beta_K)}{\partial \beta^2_1}      & \cdots & 0      \\
 \vdots & \vdots & \ddots & \vdots \\
 0      & 0      & \cdots & \frac{\partial^2 g( \beta_1,\beta_2, \cdots,\beta_K)}{\partial \beta^2_K}    \\
\end{matrix}
\right).
 \end{align}
Based on~\eqnref{equ:MIendsim24-1} and~\eqnref{equ:MIendsi31-1}, $H(\rho, \beta_1,\beta_2, \cdots,\beta_K)$ in \eqnref{equ:Hess} is a negative definite matrix~\cite{BaekP10}.
According to the criterion of extreme value of the function, when $\rho_1=\rho_2=\cdots=\rho_K=\rho^*=\Upsilon$, $\alpha_k=c_{\alpha}|\tilde{h}_k|^2$ and $\beta_k=c_{\beta}|\tilde{h}_k|^2$ for all $k$, the MI reaches the maximum.
Thus, Proposition  \ref{proposition:prop4} is proved.

\end{appendices}

\bibliographystyle{IEEEtran}

\end{document}